\documentclass[prd,preprint,superscriptaddress,amsmath,amssymb]{revtex4}

\usepackage{graphicx,color,slashed}

\begin{document}

{\begin{flushright}{KIAS-P19014}
\end{flushright}}

\title{\bf Influence of an inert charged Higgs boson on the muon $g-2$ and radiative neutrino masses in a scotogenic model  }

\author{Chuan-Hung Chen}
\email{physchen@mail.ncku.edu.tw}
\affiliation{Department of Physics, National Cheng-Kung University, Tainan 70101, Taiwan}

\author{Takaaki Nomura}
\email{nomura@kias.re.kr}
\affiliation{School of Physics, KIAS, Seoul 02455, Korea}

\date{\today}

\begin{abstract}
A simple extension of Ma's approach in a scotogenic model is studied for the purpose of simultaneously interpreting the neutrino data and the excess of muon anomalous magnetic moment (muon $g-2$). The feasible minimal extension is to add an $Z_2$-odd vector-like lepton doublet to the Ma's model. It is found that in addition to the neutrino data, the strict constraints on the relevant parameters are from the electroweak oblique parameters and the induced lepton-flavor violation processes, such as $\ell_i \to \ell_j \gamma$ and $\ell_i \to \ell^-_j \ell^-_j \ell^+_j$. Performing parameter scan,  we numerically demonstrate that when the constraint conditions are satisfied, the muon $g-2$ of  $O(10^{-9})$ can be achieved, where it  can be expected that with a $5\sigma$ observation, the Muon $g-2$ experiment at Fermilab can observe $a_\mu \approx 13.31 \times 10^{-10}$ when the current experiment and the SM errors are reduced by a factor of 4 and 2, respectively. Moreover, the branching ratio of the $\tau \to \mu \gamma$ decay can match the Belle II sensitivity of  $O(10^{-9})$ with an integrated  luminosity of   50 ab$^{-1}$. 

\end{abstract}

\maketitle

\section{Introduction}

In addition to the origin of neutrino mass, a clear hint for new physics is the muon anomalous magnetic moment (muon $g-2$). The  results measured by the E821 experiment at Brookhaven National Lab (BNL)~\cite{Bennett:2006fi} and calculated in the standard model (SM)  are respectively given as~\cite{PDG}
 \begin{align}
 a^{\rm exp}_\mu & = (11 659 209.1 \pm 5.4 \pm 3.3)\times 10^{-10}  \,, \nonumber \\
 a^{\rm SM}_\mu & = (11 659 182.3 \pm 0.1 \pm 3.4 \pm 2.6)\times 10^{-10}, 
  \end{align}
  where the uncertainties  in the SM are from the electroweak, lowest-order hadronic, and higher-order hadronic effects. The difference between the SM and experiment  is~\cite{PDG}:
   \begin{equation}
   \Delta a_\mu = a^{\rm exp}_\mu -a^{\rm SM}_\mu = (26.8 \pm 6.3 \pm 4.3)\times 10^{-10}\,,
   \end{equation}
which indicates a 3.5$\sigma$ deviation. Moreover, the recent theoretical analysis shows a 3.7$\sigma$ deviation~\cite{Keshavarzi:2018mgv}. Accordingly, resolutions to the muon $g-2$ excess have been  broadly studied in the literature~\cite{Czarnecki:2001pv,Gninenko:2001hx,Ma:2001mr,Chen:2001kn,Ma:2001md,Benbrik:2015evd,Baek:2016kud,Altmannshofer:2016oaq,Chen:2016dip,Lee:2017ekw,Chen:2017hir,Das:2017ski,Calibbi:2018rzv,Barman:2018jhz, Nomura:2016rjf,Kowalska:2017iqv}.  A detailed review of the muon $g-2$ can be found in~\cite{Jegerlehner:2009ry,Miller:2012opa,Lindner:2016bgg,Jegerlehner:2018zrj}.
 
 The new muon $g-2$ measurements performed in the E989 experiment at Fermilab and the E34 experiment at J-PARC will aim for a precision of 0.14 ppm~\cite{Grange:2015fou} and 0.10 ppm~\cite{Otani:2015jra}, in which  the experimental accuracy can be improved by a factor of $4$ and $5$, respectively. If we assume the future experimental and theoretical uncertainties can be respectively reduced by a factor of 4 and 2,  it is expected that with a 5$\sigma$ measurement, $\Delta a_\mu \approx 13.31 \times 10^{-10}$ can be observed by the Fermilab muon $g-2$ experiment, which has started taking data~\cite{Hong:2018kqx}. 

 It is a highly non-trivial issue to simultaneously generate the neutrino mass at the $10^{-2}$ eV scale and  explain the muon $g-2$ excess in a simple extension of the SM. One of feasible possibilities to accommodate both phenomena is that both processes can be achieved through the quantum radiative corrections. A known mechanism for a radiative neutrino mass of $O(10^{-2})$ eV is the  scotogenic model proposed in~\cite{Ma:2006km} (called Ma-model in this paper), where the dark matter (DM) candidate can be the lightest  inert neutral scalar or the right-handed neutrino $(N_k)$~\cite{Ma:2006km,Barbieri:2006dq}.  
 
 It is  found that the Ma-model  cannot generate a sufficient $\Delta a_\mu$ without an extension. The main reasons are as follows: (i) The lepton anomalous magnetic moment can be generated by the mediation of inert charged-Higgs and dark right-handed neutrinos. Since the involved charged leptons are left-handed, to match the chirality of tensor-type dipole operators, the effect indeed is suppressed by $m^2_{\ell}/m^2_{N_k}$. (ii) $a^{\rm NP}_\mu$ induced by a charged-Higgs at the one-loop level is usually negative~\cite{Dedes:2001nx}. Therefore, in this work, we study  whether the neutrino data and $a^{\rm NP}_\mu\sim O(10^{-9})$ can be accommodated in a scotogenic model when the Ma-model is minimally extended. 
 
 We find that the feasible minimal extension is to include an $Z_2$-odd vector-like lepton doublet ($X$). Due to the new dark lepton doublet, the left-handed and right-handed couplings  can  now appear in the same loop diagram; therefore, the induced $a^{\rm NP}_\mu$ is proportional to  $m_\mu$, not $m^2_\mu$. Because  more Yukawa couplings are involved, we have the degrees of freedom to make the inert charged-Higgs-induced $a^{\rm NP}_\mu$ positive. Although the inert neutral scalar bosons can also contribute to the muon $g-2$, due to strong cancellation and $m^2_\ell/m^2_{N_k}$ suppression, their effects are small and can be neglected. Intriguingly, it will be shown that the proposed model can originate from a larger gauge symmetry, such as $SO(10)$~\cite{Ma:2018uss,Ma:2018zuj}.

 %we find that the lightest DM prefers the lightest Majorana particle in the model. 
 Since we concentrate the study in the flavor physics, we do not analyze the DM-related physics in this study. The relevant DM analysis can be found in~\cite{Kubo:2006yx,Babu:2007sm,Gelmini:2009xd,Adulpravitchai:2009re,Schmidt:2012yg,Bouchand:2012dx,Klasen:2013jpa,Vicente:2014wga,Merle:2015gea,Merle:2015ica,Ibarra:2016dlb,Merle:2016scw,Ahriche:2016cio,Lindner:2016kqk,Borah:2017dqx, Borah:2017dfn,Hagedorn:2018spx,Bhattacharya:2018fus,Baumholzer:2018sfb,Borah:2018rca}. It is worth mentioning that it has been found that  in some parameter regions, the imposed $Z_2$ symmetry in original Ma-model could be broken when renormalization group equation (RGE) effects are taken into account~\cite{Merle:2015gea,Merle:2015ica,Lindner:2016kqk}. The possible  resolutions to the problem can be found in~\cite{Vicente:2014wga,Merle:2016scw,Ahriche:2016cio}. In addition, we also skip the analysis for the signal search at the LHC, where the related discussions can be found in Refs.~\cite{Cao:2007rm,Sierra:2008wj, Bhattacharya:2015qpa, Hessler:2016kwm,Diaz:2016udz,Boos:2018fnt,Cai:2018upp,Ahriche:2018ger,Ahriche:2017iar}.

 In addition to the neutrino physics and muon $g-2$,  lepton flavor violation (LFV) processes, such as $\ell_i \to \ell_j \gamma$ and $\ell_i \to \ell^{-}_{j} \ell^-_j \ell^+_{j}$ ($\ell_i \to 3 \ell_j$), can be produced in the extension model~\cite{Vicente:2014wga,Toma:2013zsa}.  Additionally,  $X$ and $N_k$ can together couple  through the SM Higgs doublet, so that the electroweak oblique parameters may constrain the related parameters due to the mass splitting within the vector-like lepton doublet. Hence, it is a challenge to require all related parameters through various combinations to fit the current experimental upper limits. After taking some assumptions  based on the $\mu\to e \gamma$ constraint, 11 new independent parameters are involved. We will show that the 11 free parameters can be accommodated in the model when all constraints from the electroweak oblique parameters, the LFV processes, and the neutrino data are satisfied; and the muon $g-2$ can still reach  the level of $10^{-9}$. 
 %; for instance, the branching ratios (BRs) for the radiative LFV decays are $BR(\mu \to e \gamma)< 4.2\times 10^{-3}$ and %$BR(\tau \to \mu \gamma)<4.4\times 10^{-8}$. 

When the $\mu \to e \gamma$ constraint is compromised in the model, indeed, $\tau \to \mu \gamma$ exerts an important constraint on the parameters, especially those related to the neutrino mass matrix for which we cannot arbitrarily tune the parameters  to be small.  After scanning the chosen parameter regions, it is found that the branching ratio ( BR) for the $\tau \to \mu \gamma$ decay can be well controlled in the model and that $BR(\tau \to \mu \gamma)$ can be as large as the current upper bound of $4.4\times 10^{-8}$, depending on the values of the involved parameters.   With 50 ab$^{-1}$ of data accumulated at the Belle II, the sample of $\tau$ pairs can be increased to approximately  $5 \times 10^{10}$, where the sensitivity necessary to observe the LFV $\tau$ decays can reach $10^{-10} -10^{-9}$~\cite{Aushev:2010bq}. If  Belle II observes $BR(\tau \to \mu \gamma)$ at the level of $10^{-9}$, the scotogenic model can provide the interpretation of the observation.

The paper is organized as follows: We briefly introduce the model and the relevant couplings in Sec.~II. In Sec.~III, we derive the formulas for the neutrino mass matrix, for the $\ell_i \to \ell_j \gamma$ decays, for the $\ell_i \to 3 \ell_j$ decays, and for the lepton $g-2$, respectively. Based on the neutrino oscillation data, we also show the allowed region for each neutrino mass matrix element.  
The parameter scan and the detailed numerical analysis are shown in Sec. IV. In this section, we also provide a detailed numerical analysis of the relevant phenomena.  
 A summary is given in Sec. V.

\section{Model}

In this study, we extend the SM gauge symmetry, including an  $Z_2$-parity symmetry. 
In order to generate the neutrino mass through a one-loop radiative mechanism and provide the dark matter candidate, we add three right-handed neutrinos  $N_k=(1,0)$ ($k=1,2,3$) and one inert Higgs doublet $H_I=(2,1)$ to the SM~\cite{Ma:2006km}, where both $N_k$ and $H_I$ are $Z_2$-odd states, and numbers in brackets denote the $SU(2)_L$ representation and $U(1)_Y$ hypercharge, respectively.  Using the introduced $N_k$ and $H_I$,  it is found that the muon $g-2$ can be significantly enhanced when  a vector-like lepton doublet $X_{L(R)}=(2,-1)$ is included. Since the heavy lepton doublet has to couple to the SM leptons and $Z_2$-odd particles, i.e. $N_k$ and $H_I$,  $X_{L(R)}$ must carry the  $Z_2$ charge. Thus, in addition to  $N_k$, which is free from the mixing with the SM neutrino~\cite{Ma:2006km}, in principle, the new neutral lepton $\chi^{0C}_{L}$ and scalar bosons can be the  DM candidate.  

\subsection{Yukawa couplings and mass splitting in dark lepton doublet}

The gauge invariant lepton Yukawa couplings under  $SU(2)_L \times U(1)_Y\times Z_2$ symmetry can be written as: 
 \begin{align}
 -{\cal L}_Y & = y^\ell_{ij} \bar L_i   H \ell_{R j} +  y^k_{L i} \bar L_i  \tilde H_I N_k + y_{R j}  \bar X_{L} H_I \ell_{R j}  \nonumber \\
 %+ h_R \bar X_R   \tilde{H} N^C_R  
 & +  h^k_L \bar X_L  \tilde{H} N_k
 + \frac{m_{N_k}}{2} \overline{ N^C_k} N_k + m_X \bar X_L X_R+ H.c.\,, \label{eq:Yu_D}
 \end{align}
where $i, j=1,2,3$ denote the flavor indices; $H^T=(G^+, (v+ h +i G^0)/\sqrt{2})$ is the SM Higgs doublet and $v$ is the vacuum expectation value (VEV) of $H$; $N^C = C \gamma^0 N^*$ with $C=i \gamma^0 \gamma^2$, $\tilde{H}_{(I)}=i \tau_2 H^*_{(I)}$; $m_{N}$ and $m_X$ are the masses of $N_R$ and $X_{L(R)}$, respectively, and the representations of dark $H_I$ and $X_{L(R)}$ are given as:
 \begin{equation}
 H_I = \left(
\begin{array}{c}
   H^+_I  \\
  (S_I + i A_I)/\sqrt{2}
\end{array}
\right)\,, ~~
X_{L(R)}= \left(
\begin{array}{c}
   \chi^0  \\
   \chi^-
\end{array}
\right)_{L(R)}\,.
\end{equation}
%In order to retain the lepton-number conservation in dimension-4 terms,  $\bar X_R   \tilde{H} N^C_R$, which is allowed in the %considered symmetry, has been dropped.  
 Since $\chi^-$ is an $Z_2$-odd particle and cannot mix with the SM charged leptons  after electroweak symmetry breaking (EWSB),  the SM charged-lepton masses are still dictated by the first term in Eq.~(\ref{eq:Yu_D}).  That is, the SM leptons in Eq.~(\ref{eq:Yu_D})  can be taken as the physical states after EWSB and their masses can be expressed as $m_{\ell_i}= y^{\ell}_{ij } v\delta_{ij}/\sqrt{2}$. In terms of the representation components, the new Yukawa interactions are written as:
 \begin{align}
  -{\cal L}_Y &\supset   \left( y^k_{Li} \bar \nu_{L i} N_k + y_{R i} \bar \ell_{Ri} \chi^-_L \right) \frac{S_I + i A_I}{\sqrt{2}} + h^k_L \overline{\chi^0_L} N_k  \frac{v+h}{\sqrt{2}}+ \frac{m_{N_k}}{2} \overline{N^C_k} N_k \nonumber \\
 & + \left( y_{Ri} \overline{\chi^0_L} \ell_{Ri} - y^k_{Li} \bar N_k \ell_{L i} \right)H^+_I + m_X \left(\overline{\chi^0_L} \chi^0_R + \overline{\chi^-_L} \chi^{-}_R \right) + H.c. \,, \label{eq:Yukawa}
 \end{align}
 where $y_{Ri}$ and $h^k_L$ are taken as the real parameters.

  It is worth mentioning that the proposed model can arise from a larger gauge group, such as $SO(10)$ grand unified theories (GUTs)~\cite{Ma:2018uss,Ma:2018zuj}, where the symmetry breaking chain is $SO(10)\to SU(5) \times U(1)_\chi \to SU(3)_C\times SU(2)_L \times U(1)_Y \times U(1)_\chi$. 
%Moreover we could embed such new field contents in multiplets of larger gauge group such as $SO(10)$ in grand unified theory %(GUT) where we can realize the relevant Yukawa interactions for muon $g-2$, $\bar X_L \tilde H N_k$ and $\bar X_L H_I %\ell_R$, after breaking gauge group through $SO(10) \supset SU(5) \times U(1)_\chi$~\cite{Ma:2018uss,Ma:2018zuj}.
Denoting all fermion representations as the left-handed states,   the new lepton doublets $X$ and $X^c$ can originate from ${\bf 10} $ of $SO(10)$ and can be  $({\bf \bar 5}, -2) + ({\bf 5}, 2)$ in $SU(5)\times U(1)_\chi$.
% ${\bf 10}$ of $SO(10)$ as ${\bf 10} \supset ({\bf \bar 5}, -2) + ({\bf 5}, 2)$ 
%where we denote fermion fields as left-handed and $SU(5)$ representation ${\bf 5}$ contain $SU(2)_L$ doublet as ${\bf 5} 
%\supset ({\bf 1}, {\bf 2}, 1/2)$ 
%under $SU(5) \supset SU(3)_c \times SU(2)_L \times U(1)_Y$. 
If we embed  the inert doublet $ \tilde H_I$, the right-handed neutrinos $N_k$, and the SM Higgs field  in the representations of ${\bf 16} \supset ({\bf \bar 5}, 3)$, ${\bf 45} \supset ({\bf 1}, 0)$, and ${\bf 10}$, 
%Also inert doublet $ \tilde H_I$ can appear as ${\bf 16} \supset ({\bf \bar 5}, 3)$ and right-handed neutrino $N_k$ can be 
%embedded as ${\bf 45} \supset ({\bf 1}, 0)$.
 the Yukawa interactions $\bar X_L \tilde H N_k$ and $\bar X_L H_I \ell_R$  can be gauge singlets  under the gauge symmetries ${\bf 10} \times {\bf 10} \times {\bf 45}$ and ${\bf 10} \times {\bf 16} \times {\bf 16}$, respectively.

Because the SM Higgs doublet $H$ couples to $X_{L,R}$ and $N_R$ and $\chi^0_{L,R}$ can mix with $N_R$ when the electroweak symmetry is broken, the $5\times 5$ neutral lepton mass matrix in the basis $( \chi^0_R, \chi^{0C}_L, N_k)$ can be written as:
 \begin{equation}
 M=\left(
\begin{array}{ccc} 
 0 & m_X &    {\bf 0}_{1\times 3} \\
 m_X & 0& v {\bf h}_{L}/\sqrt{2} \\
 {\bf 0}_{3\times 1} & v{\bf h}^T_L/\sqrt{2} & ({\bf m}_{N })_{3\times 3} 
\end{array}
\right) 
\,, \label{eq:matrix_mass}
 \end{equation}
 with ${\bf h_L}=(h^1_L, h^2_L, h^3_L)$ and ${\bf m}_{N }={\rm diag}(m_{N_1}, m_{N_2}, m_{N_3})$. 
 The symmetric mass matrix can be diagonalized using an orthogonal matrix.  With the assumption of $m_{N_1} = m_{N_2}=m_{N_3}=m_0$,  the eigenvalues  of the five Majorana states can be obtained  as:
 \begin{align}
m_1 &\approx  - m_X-(e_X-e_N)  \,, ~m_2 \approx  m_X +  e_X\,, ~m_{3(4)}=m_0 \,,   \nonumber \\
 m_5 & \approx m_0 - e_{N}\,, ~e_{X}  = \frac{v^2}{8 (m_X +m_{0})} \left(\eta_h\pm \sqrt{\eta^2_h + 16  \zeta_h (m_X+m_0)^2/v^2}\right)\,, \nonumber \\
 \eta_h & =\zeta_h +4 \frac{m^2_N -m^2_{X}}{v^2}\,, ~ e_N= e_X - \frac{v^2 \zeta_h}{4(m_X+m_0)} \,, \label{eq:eigenvalues}
 \end{align}
 where we define $\zeta_h= \sum_k (h^k_L)^2$, and $h_k\equiv v h^k_L/\sqrt{2}$ is taken as  the perturbative parameters, and  the $\pm$ sign in $e_{X}$ can determine what the lightest Majorana particle is, i.e., $\chi^{0C}_L$ or one of $N_{k}$.  Note that in order to  simplify the analysis for  the flavor physics, we set all $m_{N_k}$ to be the same although generally this is not necessary. If the DM candidate is the lightest  right-handed neutrino ($N_{k'}$), we can  take $m_{N_{k'}}$ to be smaller than the others. Since our main target is on the  flavor physics,  we  do not further pursue the DM issue in this work. The relevant discussion can be found in~\cite{Ma:2006km,Kubo:2006yx,Babu:2007sm,Gelmini:2009xd,Adulpravitchai:2009re,Schmidt:2012yg,Bouchand:2012dx,Klasen:2013jpa,Vicente:2014wga,Merle:2015gea,Merle:2015ica,Ibarra:2016dlb,Merle:2016scw,Ahriche:2016cio,Lindner:2016kqk,Borah:2017dqx, Borah:2017dfn,Hagedorn:2018spx,Bhattacharya:2018fus}. Using the obtained eigenvalues, the flavor mixing matrix can be approximately formulated as:
  \begin{equation}
   {\cal O}_X \approx
\left(
\begin{array}{ccccc}
\frac{m_X}{N_{-X}|m_1|} &  -\frac{1}{N_{-X}} &  \frac{h_1}{2m_0N_{-X}} &  \frac{h_2}{2m_0N_{-X}} &  \frac{h_3}{2m_0N_{-X}} \\
\frac{m_X}{N_Xm_2} &  \frac{1}{N_{X}}  &  -\frac{h_1}{N_X (m_0 -m_2)} &   -\frac{h_2}{N_X (m_0 -m_2)} &  -\frac{h_3}{N_X (m_0-m_2)} \\
 0 & 0 & \frac{h_2}{N_{N_1} \sqrt{h^2_1 + h^2_2} }&  - \frac{h_1}{N_{N_1} \sqrt{h^2_1 + h^2_2} }  & 0\\
  0 & 0 & \frac{h_1}{N_{N_2} \sqrt{h^2_1 + h^2_2} }&   \frac{h_2}{N_{N_2} \sqrt{h^2_1 + h^2_2} }  & - \frac{\sqrt{h^2_1 + h^2_2}}{N_{N_2} h_3}\\
 \frac{m_X}{N_{N_3}m_5} &   \frac{1}{N_{N_3}}&  -\frac{h_1}{N_{N_3} (m_0 -m_5)}  & -\frac{h_2}{N_{N_3} (m_0-m_5)} & \frac{h_3}{N_{N_3} (m_0-m_5)}
\end{array}
\right)\,,
   \end{equation}
where  $N_{a}$ ($a=-X,X,N_k$) are the normalization factors, which follow $\sum_i {\cal O}_{X a i}^2 =1$. 

From the results,  it can be seen that the mass splitting within the vector-like lepton doublet can be expressed as $\Delta m_X =  |m_2 - m_X| \approx |e_X|$ and that it depends on $v h^k_L/\sqrt{2}$.
This mass splitting contributes to the electroweak oblique parameters, where the current measurements with $U=0$ are given as~\cite{PDG}:
\begin{equation}
S = 0.07 \pm 0.08\,, ~ T =0.10 \pm 0.07\,. \label{eq:ST_data}
 \end{equation}
Therefore, the precision measurements of electroweak oblique parameters~\cite{Peskin:1991sw} may constrain $h^k_L$.  In order to consider the constraints, we write the oblique corrections for the vector-like lepton doublet as~\cite{Lavoura:1992np,Arina:2012aj}:
 \begin{align}
  S &= \frac{1}{\pi} \left[ \frac{22 z_1 + 14 z_2}{9}- \frac{1}{9} \ln\frac{z_1}{z_2} + \frac{11z_1 +1}{18} f(z_1) + \frac{7 z_2 -1}{18} f(z_2)   \right.\nonumber \\
  & \left. ~~~ - \sqrt{z_1 z_2} \left( 4 + \frac{f(z_1)+ f(z_2)}{2}\right)\right]\,, \nonumber \\
 T &= \frac{1}{8 \pi s^2_W c^2_W} \left[ z_1 + z_2 - \frac{2 z_1 z_2}{z_1 - z_2} \ln \frac{z_1}{z_2} + 2 \sqrt{z_1 z_2} \left( \frac{z_1+z_2}{z_1 - z_2} \ln \frac{z_1}{z_2}-2\right)\right]\,,  \label{eq:ST}\\
 f(x)&=-4 \sqrt{4 x-1} \, {\rm arctan}\frac{1}{\sqrt{4x-1}}\,,\nonumber
 \end{align}
with $z_1=(m_0 - e_X)^2/m^2_Z$ and $z_2=m^2_0/m^2_Z$. Since the $U$ parameter usually is small, we do not explicitly  show it. 

 In the calculations of LFV processes, we need the gauge couplings to the photon and $Z$-gauge boson. The relevant interactions are given as:
  \begin{align}
  {\cal L}_{V} & = - Q_\ell e \bar \ell  \gamma_\mu \ell A^\mu - i e (H^-_I \partial_\mu  H^+_I -  \partial_\mu (H^-_I) H^+_I) A^\mu - \frac{g}{2\cos\theta_W} \overline{\chi^0} \gamma_\mu \chi^0 Z^\mu \nonumber \\ & - \bar \ell \gamma_\mu  ( C^\ell_L P_L + C^\ell_R P_R ) \ell Z^\mu - i  \frac{g\cos2\theta_W}{2\cos\theta_W} (H^-_I \partial_\mu  H^+_I - H^+_I \partial_\mu H^-_I ) Z^\mu\,, \label{eq:gauge_c}
  \end{align}
with
 \begin{equation}
 C^\ell_L = \frac{g}{2\cos\theta_W} (-1 + 2 \sin^2\theta_W)\,, ~ C^\ell_R = \frac{ g \sin^2\theta_{W}}{\cos\theta_W}\,.
 \end{equation}

\subsection{Scalar potential and gauge couplings to dark sector} 

The gauge invariant  scalar potential with the $Z_2$-parity can be written as~\cite{Ma:2006km,Barbieri:2006dq}:
\begin{align}
V(H,H_I) &= \mu^2 H^\dagger H + \lambda_1 ( H^\dagger H)^2 + m^2_{I}  H^\dagger_I H_I + \lambda_2 ( H^\dagger_I H_I )^2\nonumber \\
& +\lambda_3 H^\dagger H H^\dagger_I H_I + \lambda_4 H^\dagger H_I H^\dagger_I H + \frac{\lambda_5}{2}\left[ (H^\dagger H_I)^2 +H.c. \right] \,, \label{eq:V}
\end{align}
where  $v=\sqrt{-\mu^2/\lambda_1}$ with $\mu^2 <0$ and $m_h=\sqrt{v^2 \lambda_1/2} $ are the same as the SM, and the massive inert Higgs doublet  requires $m^2_I >0$. With $v\approx 246$ GeV and $m_h\approx 125$ GeV, we can obtain $\lambda_1 \approx 0.516$. 
The masses of $(S, A, H^\pm)$  can be expressed as~\cite{Ma:2006km,Barbieri:2006dq} :
\begin{equation}
m^2_{S_I}= m^2_I + \lambda_L v^2\,, ~m^2_{A_I} - m^2_{S_I} = - \lambda_5 v^2\,, ~m^2_{H^\pm_I}=m^2_I + \frac{\lambda_3} {2}v^2
\end{equation}
with $\lambda_L=(\lambda_3+\lambda_4+\lambda_5)/2$. It can be seen that  the mass difference between $S_I$ and $A_I$ is dictated by the $\lambda_5$ parameter. We will show that in addition to the Yukawa couplings, the radiative neutrino mass also depends on the mass difference. If $y^k_{Li} \sim O(10^{-2})$ are required,  $|\lambda^{5}|\sim 10^{-8}$ is necessary to fit the neutrino mass matrix elements, which are  of the $O(10^{-2})$ eV. 

In the model, the DM particle can be the lightest $N_{k}$ or $S_{I}(A_{I})$. If we select $S_I$ or $A_{I}$ as the DM candidate, in order to escape the constraint from the DM-nucleus scattering, which is generated  through the $S_I A_I Z$ gauge coupling~\cite{Barbieri:2006dq}, $|m_{A_I}-m_{S_I}|$ must have a low limit in order to kinematically suppress the scattering process. Then, $y^k_{Li}$ have to be of the order of $10^{-4}-10^{-3}$ to match the neutrino mass matrix elements. As a result, the muon $g-2$ arising from the inert charged-Higgs is suppressed. Similarly, $\chi^0$ cannot be the DM candidate because the gauge coupling $\chi^0 \chi^0 Z$ leads a large cross section in the process of DM scattering off the  nucleus.   
 Hence, we will concentrate on the case with $m_{S_I(A_I), H^\pm_I} > m_{0,X}$. 

\section{Radiative neutrino mass,  LFV,  and lepton $g-2$ }

 In this section, we derive the formulas for the neutrino mass matrix, the $\ell_i \to \ell_j \gamma$ and $\ell_i \to 3\ell_j$ processes, and lepton $g-2$ in the model. Although the original Ma's model can provide sizable contributions to the LFV processes, we  checked that with $y^{k*}_{1} y^k_{2} \sim O(10^{-3})$, the BR for $\mu \to e \gamma$ is of the order of $10^{-15}$, which is two orders of magnitude smaller than the current upper limit. Therefore, in the following analysis, we concentrate on the extension effects. 

\subsection{Radiative neutrino mass}

\begin{figure}[phtb]
\includegraphics[scale=0.75]{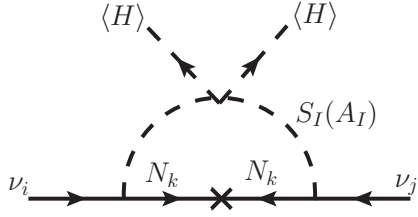}
 \caption{  Feynman diagram for radiative neutrino mass.  }
\label{fig:neutrino_loop}
\end{figure}

The  Majorana neutrino mass arisen from a quantum loop in the scotogenic model is sketched in Fig.~\ref{fig:neutrino_loop}. It can be seen that  in addition to the Yukawa couplings, the essential effect is from the $(H^\dagger H_I)^2$ coupling, which is dictated by the $\lambda_5$ parameter.  From the couplings in Eq.~(\ref{eq:Yukawa}) and Eq.~(\ref{eq:V}), 
 the Majorana neutrino mass matrix elements can be  obtained as~\cite{Ma:2006km,Cai:2017jrq}:
 \begin{equation}
 m^{\nu}_{ij}  = \sum_k \frac{ y^k_{Li} y^{k}_{Lj}}{2(4\pi)^2} m_{N_k} \left[  \frac{ m^2_{A_I} \ln(m^2_{A_I}/m^2_{N_k})} {m^2_{N_k} -m^2_{A_I}}- \frac{ m^2_{S_I} \ln(m^2_{S_I}/m^2_{N_k})} {m^2_{N_k} -m^2_{S_I}}  \right]\,. \label{eq:mnu_ij}
 \end{equation}
It can be found that $m^{\nu}_{ij}$ can be of the $O(10^{-2})$ eV when $\sum_k y^k_{Li} y^{k}_{Lj}\sim O(10^{-4}-10^{-3})$, and $m_{S_I(A_I)}\approx m_{N_k}\approx 1$ TeV are used.  

%Moreover, in order to suppress the  DM scattering  off nucleus, which is through the $S_I A_I Z$ gauge coupling~\cite{Barbieri:2006dq}, %in the following analysis, we take $m_{A_I}-m_{S_I} \approx 0.4-1$ MeV because the typical recoil energy of nucleus in the detector is %of $O(10-100)$ keV~\cite{Undagoitia:2015gya}.

     %%%%

 The mass matrix can be diagonalized by the Pontecorvo-Maki-Nakagawa-Sakata (PMNS) matrix as:
 %The mass matrix is related to MNS matrix $U_{MNS}$ and diagonalized one $m_\nu^{\rm diag}$ by
 \begin{equation}
 m^{\nu}_{ij} = U^*_{MNS} m_\nu^{\rm diag} U^\dagger_{MNS},
 \end{equation}
 where $m_\nu^{\rm diag} = {\rm diag}(m_1, m_2, m_3)$, and 
 the PMNS matrix can be parametrized as~\cite{PDG}:
 \begin{align}
\label{MNS}
U_{\rm MNS} & = \begin{pmatrix} 
c_{12} c_{13} & s_{12} c_{13} & s_{13} e^{-i \delta} \\
-s_{12} c_{23} - c_{12} s_{23} s_{13} e^{i \delta} & c_{12} c_{23} -s_{12} s_{23} s_{13} e^{i \delta} & s_{23} c_{13} \\
s_{12} s_{23} - c_{12} c_{23} s_{13} e^{i \delta} & -c_{12} s_{23} - s_{12} c_{23} s_{13} e^{i \delta} & c_{23} c_{13} 
\end{pmatrix} \nonumber \\
& \times \text{diag}(1, e^{i \alpha_{21}/2}, e^{i\alpha_{31}/2})\,,
\end{align}
in which  $s_{ij} \equiv \sin \theta_{ij}$, $c_{ij} \equiv \cos \theta_{ij}$; $\delta$ is the Dirac CP violating phase, and $\alpha_{21, 31}$ are  Majorana CP violating phases.  Since the mass ordering is still uncertain, the current neutrino data can be shown in terms of the different mass ordering as~\cite{PDG}:
 \begin{align}
\Delta m^2_{21} & = (7.53\pm0.18)\times 10^{-5} ~{\rm eV^2}\,, ~ \sin^2\theta_{12}=0.307 \pm 0.013 \,, \nonumber \\
\Delta m^2_{32} &= ( 2.51 \pm 0.05, \, -2.56 \pm 0.04)\times 10^{-3} ~{\rm eV^2} ~ (\rm {NO,\, IO})\,, \nonumber \\
\sin^2\theta_{23} &= ( 0.597^{+0.024}_{-0.030},\, 0.592^{+0.023}_{-0.030} )  ~ (\rm {NO,\, IO})\,, \nonumber \\
\sin^2\theta_{13} & = (2.12 \pm 0.08)\times 10^{-2}\,,
 \end{align}
where $\Delta m^2_{ij}\equiv m^2_i - m^2_j$, and $\Delta m^2_{32}>0$ and $\Delta m^2_{32} < 0$ denote the normal ordering (NO) and inverted ordering (IO), respectively.  

Based on the neutrino oscillation data,  the central values of $\theta_{ij}$, $\delta$, and $\Delta m^2_{ij}$ using the global fit can then be obtained as~\cite{deSalas:2017kay}:
\begin{align}
 {\rm NO} :\, &  \theta_{12} = 34.5^{\circ}\,, \,\theta_{23}=47.7^{\circ}\,, \,\theta_{13} =8.45^{\circ}\,, \, \delta=218^{\circ}\,,\nonumber \\
&  \Delta m^2_{21} =7.55 \times 10^{-5}\, {\rm eV^2}\,, \, \Delta m^2_{31} = 2.50\times 10^{-3}\, {\rm eV^2}\,, \nonumber \\
 {\rm IO} :\, &  \theta_{12} = 34.5^{\circ}\,,\,\theta_{23}=47.9^{\circ}\,,\,\theta_{13} =8.53^{\circ}\,, \, \delta=281^{\circ}\,, \nonumber \\
 &  \Delta m^2_{21} =7.55 \times 10^{-5}\, {\rm eV^2}\,, \, \Delta m^2_{31} =- 2.42\times 10^{-3}\, {\rm eV^2}\,, 
  \end{align}
  where $m_{1(3)}=0$ for NO (IO) are applied, and the Majorana phases are taken to be $\alpha_{21(31)}=0$. 
 Taking the $3\sigma$ uncertainties, the magnitudes of the Majorana matrix elements  in units of eV for NO and IO can be respectively estimated as:
\begin{align}
\begin{pmatrix} |m^\nu_{11}| & |m^\nu_{12}| & |m^\nu_{13}| \\ |m^\nu_{21}| & |m^\nu_{22}| & |m^\nu_{23}| \\ |m^\nu_{31}| & |m^\nu_{32}| & |m^\nu_{33}| \end{pmatrix}_{\rm NO} & 
\simeq \begin{pmatrix} 0.11-0.45 & 0.12-0.82 & 0.12-0.82 \\  0.12-0.82 & 2.4-3.3 & 2.0-2.2 \\ 0.12-0.82 & 2.0-2.2  & 2.2-3.1\end{pmatrix} \times 10^{-2}\,, \nonumber \\
\begin{pmatrix} |m^\nu_{11}| & |m^\nu_{12}| & |m^\nu_{13}| \\ |m^\nu_{21}| & |m^\nu_{22}| & |m^\nu_{23}| \\ |m^\nu_{31}| & |m^\nu_{32}| & |m^\nu_{33}| \end{pmatrix}_{\rm IO} & 
\simeq \begin{pmatrix} 4.8-5.0 & 0.41-0.65 & 0.39-0.62 \\  0.41-0.65 & 1.9-2.8 & 2.4-2.6 \\ 0.39-0.62 & 2.4-2.6 & 2.2-3.1\end{pmatrix} \times 10^{-2}
\,. \label{eq:v_nu_mass}
\end{align}
It can be found that  when $\sum _k y^k_{Li' } y^{k}_{Lj'} \sim  10^{-3}$ ($i',j'=2,3$) and 
\begin{equation}
{\cal M}_0 = \frac{m_{0}}{16 \pi^2}  \left[ \frac{ m^2_{A_I} \ln(m^2_{A_I}/m^2_{0})} {m^2_{0} -m^2_{A_I}} - \frac{ m^2_{S_I} \ln(m^2_{S_I}/m^2_{0})} {m^2_{0} -m^2_{S_I}} \right] \sim   \ {\rm eV}, 
\end{equation}
  $m^\nu_{i', j'} \sim O(10^{-2})$ eV can then be obtained.  We will show that due to the $\mu\to e\gamma$ constraint,  the $y^k_{L1}$-related parameters have to be smaller than $y^k_{L2, L3}$; therefore, $m^\nu_{1i}$ are preferred to be smaller than $m^\nu_{i' j'}$, i.e. the model is suitable for the NO case.

  \subsection{ Radiative $\ell \to \ell' \gamma$ decays}
 
 In the  model, the LFV processes can arise from the  $S_I$, $A_I$, and $H^\pm_I$ boson exchanges. Since $m_{S_I}\approx m_{A_I}$ is taken in this work,   the $S_I$- and $A_I$-induced LFV effects have  strong cancellations. Thus, in this study, we concentrate  on the inert charged-Higgs effects.  The current  experimental upper limits on the BR for the relevant LFV processes are shown in Table~\ref{tab:V_LFV}.
 
 \begin{table}[htp]
\caption{Current experimental upper limits on the LFV processes. }
\begin{tabular}{c|cccc} \hline \hline
 LFV & $\mu \to e \gamma$ & $\mu \to 3 e$  &  $\tau \to \mu (e) \gamma$ & $\tau \to 3 \mu (3 e)$ \\ \hline 
 BR &  ~ $4.2 \times 10^{-13}$ ~&~  $1.0 \times 10^{-12}$ ~&~ $4.4(3.3) \times 10^{-8}$ ~&~ $2.1(2.7) \times 10^{-8}$ \\ \hline \hline
\end{tabular}
\label{tab:V_LFV}
\end{table}%
 
\begin{figure}[phtb]
\includegraphics[scale=0.8]{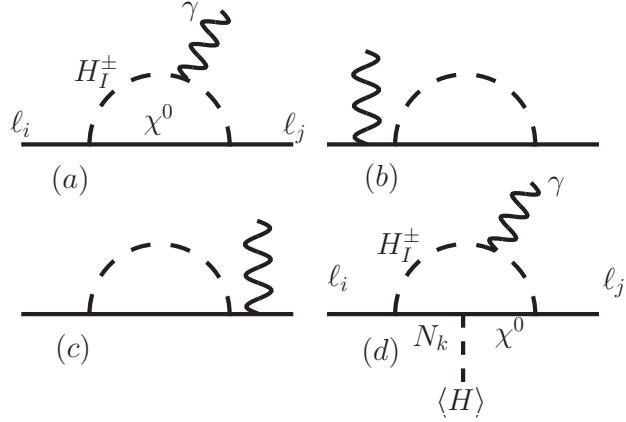}
 \caption{  $\ell_i \to \ell_j \gamma$ mediated by the inert charged-Higgs and the $\chi^0$ fermion, where plot (d) includes the mixing effect between $N_k$ and $\chi^0$. }
\label{fig:gamma_Pen}
\end{figure}

 The Feynman diagrams for the $H^\pm_I$-mediated radiative $\ell_i \to \ell_j \gamma$ decays are sketched in Fig.~\ref{fig:gamma_Pen}, where the plot (a) arises from the $H^\pm_I$ and $\chi^0$-fermion loop and the plots (b) and (c) are the associated self-energy diagrams, which can be used to remove the ultraviolet divergence.   According to the Yukawa couplings in Eq.~(\ref{eq:Yukawa}), the effective interactions for $\ell_i\to \ell_j\gamma$ can then be obtained as:
  \begin{equation}
-  {\cal L}^{\chi^0}_{\ell_i \to \ell_j \gamma}  =  - \frac{e}{2} m_{\ell_i} a^{ji}_L \bar \ell_j  \sigma_{\mu \nu}   P_L  \ell_i F^{\mu \nu}\,, \label{eq:int_gamma}
 \end{equation}
 where the Wilson coefficient and loop integral are given as:
 \begin{align}
a^{ji}_{L}  =  \frac{ y^*_{Rj} y_{Ri} }{16 \pi^2 m^2_X} I^\gamma_{L}\left(\frac{m^2_{H^\pm_I}}{m^2_X}\right)\,, \label{eq:LFV_as}  \\
 I^\gamma_{L}(a)=\int^1_0 dx \int^x_0 dy \frac{(x-1)(x-y)}{1-x + a x}\,. \nonumber 
 \end{align}
 Because $m_{\ell_j} \ll m_{\ell_i}$, we have neglected the $m_{\ell_j}$ effects. Since only right-handed leptons couple to $\chi^0$, in order to match the chirality of the dipole operator,  a mass insertion in the $\ell_i$ leg to flip the $\ell_i$ chirality from the right-handed state to the left-handed state becomes necessary. As a result, Eq.~(\ref{eq:int_gamma}) is proportional to $m_{\ell_i}$, and the left-handed $\ell_i$ is involved in the radiative decay.  We note that although $\ell_i \to \ell_j \gamma$ processes can be induced through the $N_k$ mediators, because the associated Yukawa couplings $y^{k*}_{Lj} y^k_{Li}$ are constrained by the  neutrino masses to be of $O(10^{-4}-10^{-3})$~\cite{Ma:2006km}, we thus neglect their contributions.  

    %%%%

In addition to the $H^\pm_{I}$-$\chi^0$ loops shown in Figs.~\ref{fig:gamma_Pen}(a)-(c), the $\ell_i \to \ell_j \gamma$ can be generated by Fig.~\ref{fig:gamma_Pen}(d), where  the diagram involves the  mixing of $\chi^0$ and  $N_k$, where the mixing occurs through the VEV of the SM Higgs field, i.e. $v h^k_L/\sqrt{2}$. Because $N_k$ and $\chi^0$ have been massive particles before EWSB, it is more convenient to use the weak  eigenstates of $N_k$ and $\chi^0$ to estimate Fig.~\ref{fig:gamma_Pen}(d).  Accordingly, the  effective interactions for $\ell_i \to \ell_j \gamma$ can be written as:
 \begin{equation}
 -{\cal L}^{N \chi^0}_{\ell_i \to \ell_j \gamma}  =  - \frac{e}{2} m_{\ell_i}\bar \ell_j  \sigma_{\mu \nu}  \left( b^{ji}_L P_L + b^{ji}_R P_R \right) \ell_i F^{\mu \nu}\,, \label{eq:gamma_Nchi}
 \end{equation}
where the Wilson coefficients are obtained as:
 \begin{align}
 b^{ji}_L &= \frac{  1 }{16 \pi^2 }   \frac{v  y^*_{Rj} \xi_{Li} }{\sqrt{2} m_{\ell_i} m^2_X}I^\gamma_{N_k \chi^0}\left(\frac{m^2_{N_k}}{m^2_X},\frac{m^2_{H^\pm_I}}{m^2_X}\right) \,,\nonumber \\
  b^{ji}_R &= \frac{1}{16 \pi^2 }  \frac{v  \xi^*_{Lj} y_{Ri}}{\sqrt{2} m_{\ell_i} m^2_X}I^\gamma_{N_k \chi^0}\left(\frac{m^2_{N_k}}{m^2_X},\frac{m^2_{H^\pm_I}}{m^2_X}\right)\,, \label{eq:LFV_bs} \\
  I^\gamma_{N \chi^0}(a,b)& = \int^1_0 dx \int^{x}_0 dy \frac{2(1-y)y -y^2}{1-(1-a)x -(a-b) y}\,. \nonumber 
 \end{align}
 Since the  $h^k_L$ parameters always appear to be associated with $y^k_{Li}$, we define the independent $\xi_{Li}=y^k_{Li} h^k_{L}$ parameters to combine the $h^k_L$ and $y^k_{Li}$ effects. In the numerical analysis, we take all $m_{N_k}$ to be the same; therefore, $y^k_{Li} h^k_{L}$ can be read as the sum of all $k$. 
Because the left- and right-handed lepton couplings  appear in Fig.~\ref{fig:gamma_Pen}(d) at the same time, it can be seen that  the mass insertion in the $\ell_i$ leg is not  necessary. In order to combine this effect with that arisen from the $\chi^0$ loop, the $m_{\ell_i}$ factor is shown in Eq.~(\ref{eq:gamma_Nchi});  as a result,  $b^{ji}_{L,R}$ are  $1/m_{\ell_i}$-dependent. 
 Combining Eqs.~(\ref{eq:int_gamma}) and (\ref{eq:gamma_Nchi}), the BR for $\ell_i \to \ell_j \gamma$ can be expressed as
 
 \begin{equation}
 BR(\ell_i \to \ell_j \gamma) = \tau_{\ell_i} \frac{\alpha m^5_{\ell_i}}{4}  \left( |T^{ji} _L|^2 + |T^{ji}_R |^2\right)\,,\label{eq:BR_LFV_gamma}
  \end{equation}
  with $\alpha=e^2/4\pi$, and 
  \begin{equation}
  T^{ji}_L = a^{ji}_L + b^{ji}_L\,,~~~T^{ji}_R = b^{ji}_R\,.  
  \end{equation}
 
 \subsection{$\ell_i \to 3 \ell_j$ decays}
 
 The $\ell_i \to 3 \ell_j$ decays in the model can arise from the photon-penguin diagrams, e.g.  Fig.~\ref{fig:gamma_Pen} with off-shell photon, the $Z$-penguin diagrams, and the box diagrams. We show each decay amplitude as follows: For the photon-penguin diagrams,  we write the decay amplitude as:
 \begin{align}
M(\ell_i \to 3 \ell_j)^\gamma & = e^2 \bar u_j(k_1) \left[ C^{\gamma ji}_R  \gamma_\mu P_R + \frac{m_{\ell_i}}{k^2} i \sigma_{\mu \nu} k^\nu (T^{ji}_L P_L + T^{ji}_R P_R ) \right] u_i(p) \nonumber \\
& \times   \bar u_j (k_2) \gamma^\mu v_j(k_3) - \left( k_1 \leftrightarrow k_2 \right)
 \end{align}
 where $C^{\gamma ji}_R$  from  Fig.~\ref{fig:gamma_Pen}(a) is given as:
 \begin{align}
 C^{\gamma ij}_{R} & = \frac{y_{Ri} y^*_{Rj}}{16 \pi^2 m^2_X} I_1\left(\frac{m^2_{H^\pm_I}}{m^2_X}\right)\,, ~ I_1(a)  =  \int^1_0 d x \frac{x}{1-x+a x}\,.
 \end{align}
 Although Fig.~\ref{fig:gamma_Pen}(d) can also generate vectorial current-current interaction, since its numerical contribution is at least one order of magnitude  smaller than $C^{\gamma ji}_R$, we have ignored its contribution.  Using the results in~\cite{Arganda:2005ji,Hisano:1995cp}, the BR for $\ell_i \to 3 \ell_j$ induced by the photon-penguin can be expressed as:
 \begin{align}
BR(\ell_i \to 3 \ell_j)^\gamma & = \tau_{\ell_i}\frac{\alpha^2 m^5_{\ell_i} }{32 \pi} \left[ |C^{\gamma ji}_{R}|^2 - 4 Re(C^{\gamma ji}_{R} T^{ji*}_L ) \right. \nonumber \\
 & \left. + \left(|T^{ji}_L|^2 + |T^{ji}_R|^2\right) \left( \frac{16}{3} \ln\frac{m_{\ell_i}}{m_{\ell_j}} -\frac{22}{3}\right)
 \right]\,,
 \end{align}
 where $\tau_{\ell_i}$ denotes the $\ell_i$ lifetime. 

The lepton-flavor changing $\ell_i \to \ell_j Z$ can be induced by  the $Z$-penguin diagrams. In addition to being  the same diagrams  shown  in  Fig.~\ref{fig:gamma_Pen} but using the $Z$-boson instead of the photon, the $Z$-boson can also be emitted from $\chi^0$, as shown in Figs.~\ref{fig:gamma_Pen}(a) and (d). It is found that the $\ell_i \to \ell_j Z$  decays arisen from Figs.~\ref{fig:gamma_Pen}(a)-(c)  are suppressed by $m_{\ell_i} m_{\ell_j}/m^2_X$, where the same results are also shown in the $N_k$ fermion loop obtained in ~\cite{Toma:2013zsa}. If we apply the approximation with $m_{\ell_j} \approx 0$,  their contributions can be neglected. Thus, the dominant effects indeed are from Fig.~\ref{fig:gamma_Pen}(d)-related  diagrams, and the induced effective interactions can be written as:
\begin{equation}
-{\cal L}^{N\chi^0}_{\ell_i \to \ell_j Z} = \bar \ell_j \gamma_\mu \left( C^{Z ji}_L P_L + C^{Z ji}_R P_R \right) \ell_i Z^\mu\,,
\end{equation}
where the $C^{Z ji}_{L,R}$ coefficients are given as:
\begin{align}
C^{Z ji}_L & = \frac{\xi^*_{Lj}  y_{Ri} }{16\pi^2} \frac{g}{2\cos\theta_W}\frac{ v m_{\ell_i} }{\sqrt{2} m^2_X} I^Z_{N\chi^0}\left(\frac{m^2_{N_k}}{m^2_X},\frac{m^2_{H^\pm_I}}{m^2_X}\right)\,, \nonumber \\
C^{Z ji}_R & = \frac{ y^*_{Rj} \xi_{Li}}{16\pi^2}  \frac{g}{2\cos\theta_W}\frac{ v  m_{\ell_i} }{\sqrt{2} m^2_X} I^Z_{N\chi^0}\left(\frac{m^2_{N_k}}{m^2_X},\frac{m^2_{H^\pm_I}}{m^2_X}\right)\,, \nonumber \\
I^Z_{N\chi^0}(a,b) & = \int^1_0 dx \int^x_0 dy \int^y_0 dz \frac{z}{\left[ 1-(1-a)x - (a-b)z \right]^2}\,.
\end{align}
 From the result, the decay amplitude for  $\ell_i \to 3 \ell_j$  through the $Z$-penguin can be expressed as:
 \begin{align}
 M(\ell_i \to 3\ell_j)^Z &= \frac{1}{m^2_Z}  \bar u_j(k_1) \gamma_\mu \left( C^{Z ji}_L P_L + C^{Z ji}_R P_R \right) u_i(p) \nonumber \\
 &  \times \bar u_j(k_2) \gamma^\mu \left( C^\ell_L P_L + C^\ell_R P_R \right) v_j(k_3) - \left( k_1 \leftrightarrow k_2 \right)\,.
 \end{align}
Accordingly, the BR for $\ell_i \to 3 \ell_j$ can be obtained as~\cite{Arganda:2005ji,Hisano:1995cp}:
\begin{equation}
BR(\ell_i \to 3 \ell_j)^{Z}  =  \tau_{\ell_i}\frac{\alpha^2 m^5_{\ell_i} }{32 \pi} \left[ \frac{2}{3} \left( |F_{LL}|^2 + |F_{RR}|^2\right) + \frac{1}{3}\left(|F_{LR}|^2 + |F_{RL}|^2\right)\right]\,,
\end{equation}
where $F_{LL,RR}$ and  $F_{LR,RL}$ are defined as:
 \begin{align}
 F_{LL} & =\frac{C^{Z ji}_L C^{\ell}_{L}}{g^2 \sin^2\theta_W m^2_Z}\,, ~~  F_{RR} =\frac{C^{Z ji}_R C^{\ell}_{R}}{g^2 \sin^2\theta_W m^2_Z}\,, \nonumber \\
  F_{LR} & =\frac{C^{Z ji}_L C^{\ell}_{R}}{g^2 \sin^2\theta_W m^2_Z}\,,~~  F_{RL}  =\frac{C^{Z ji}_R C^{\ell}_{L}}{g^2 \sin^2\theta_W m^2_Z}\,.
 \end{align}

\begin{figure}[phtb]
\includegraphics[scale=0.7]{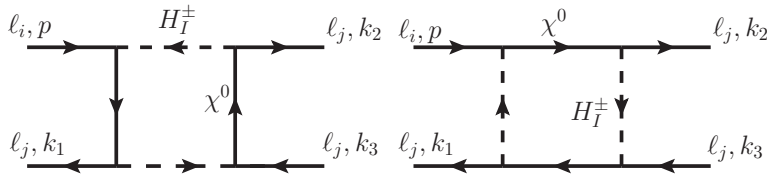}
 \caption{ Box diagrams mediated by $H^\pm_I$ and $\chi^0$ for the $\ell_i \to 3 \ell_j$ decays.   }
\label{fig:box_3l}
\end{figure}

The box diagrams mediated by $H^\pm_I$ and $\chi^0$ for $\ell_i \to 3 \ell_j$ are shown in Fig.~\ref{fig:box_3l}. Although the box diagrams mediated by $H^\pm_I$ and $N_k$  can also contribute to the $\ell_i \to 3 \ell_j$ decays, because the involving couplings are constrained by the neutrino masses, their effects can be neglected. In addition,  there are strong cancellations  between the $S_I$-$S_I$ ($A_I$-$A_I$) and $S_I(A_I)$-$A_I(S_I)$ box diagrams, so we also ignore the inert scalar and pseudoscalar  contributions. Hence, the decay amplitude for $\ell_i \to 3 \ell_j$ from the Fig.~\ref{fig:box_3l} can be obtained as:
\begin{align}
M(\ell_i \to 3\ell_j)^{\rm Box} & = C^{Bji}_{R} \bar u_j(k_1) \gamma_\mu P_R u_i(p) \, \bar u_j(k_2) \gamma^\mu P_R v_i(k_3)\,,  \\
C^{B ji}_R & =  \frac{1 }{ 16 \pi^2 m^2_X} y^*_{Rj} y_{Rj} y^*_{Rj} y_{Ri} I_{B}\left( \frac{m^2_{H^\pm_I}}{m^2_X}\right)\,, \nonumber \\
I_B(a) & = \int^1_0 dx \frac{ x(1-x)}{1-x + a x}\,. \nonumber 
\end{align}
The  BR can be found as~\cite{Arganda:2005ji,Hisano:1995cp}:
\begin{align}
BR(\ell_i \to 3 \ell_j)^{\rm Box}  =  \tau_{\ell_i}\frac{ m^5_{\ell_i} }{512 \pi^3} \frac{ |C^{B ji}_R|^2}{6}\,,
\end{align}

    %%%%

\subsection{Lepton anomalous magnetic dipole moment }

It is known that the lepton $g-2$ originates from the radiative quantum corrections, where   the form factors associated with the quantum effects can be  written as:
 \begin{equation}
 \Gamma^\mu = \bar \ell (p') \left[ \gamma^\mu  F_1(k^2) + \frac{i \sigma^{\mu \nu} k_\nu}{2m_\ell} F_2(k^2) \right] \ell(p) \,.
 \end{equation}
 The lepton $g-2$ can then be defined as:
  \begin{equation}
  a_\ell = \frac{g_\ell-2}{2} = F_2(0)\,.
  \end{equation}
 Using this definition, it can be seen that the lepton $g-2$ induced by Figs.~\ref{fig:gamma_Pen}(a)-(c) are suppressed by $m^2_\ell/m^2_X$, whereas $a_\ell$ generated by Fig.~\ref{fig:gamma_Pen}(d) is dictated by $m_{\ell}/m_X \cdot v h^k_L/m_X$. Thus, the dominant lepton $g-2$ in the model can be obtained as:
\begin{align}
a_\ell & = - \frac{ Re( \xi^*_{L \ell}  y_{R
\ell} )}{16 \pi^2} \frac{\sqrt{2} m_\ell v }{m^2_{X}}  \; I^\gamma_{N_k \chi^0}\left(\frac{m^2_{N_k}}{m^2_X},\frac{m^2_{H^\pm_I}}{m^2_X}\right) \,. \label{eq:lepton_gm2}
\end{align}
Although $a_\mu$ is associated with  $\sum_k y^k_{L2}h^k_{L}$, which are related to the neutrino masses, the muon $g-2$ can be still enhanced to $10^{-9}$ if $\xi_{L2} \sim O(0.01)$ and $y_{R2} \sim O(1)$  are allowed. In order to satisfy the strict constraints from the $\mu\to \ell \gamma$ and $\mu \to 3e$ decays,  we can take the related Yukawa couplings, e.g. $\xi_{L1}$ and $y_{R1}$ to be small. Then, the electron $g-2$ is far below the current experimental accuracy in the model. 

Before analyzing the relevant phenomena in detail, we roughly estimate the BRs for  $\ell_i \to \ell_j \gamma$ and the BRs for $\ell_i \to 3\ell_j$, which individually arise from  the photon-penguin, $Z$-penguin, and box diagrams. For illustration,  a benchmark for the relevant parameters is taken as follows:
 \begin{align}
 \xi_{L1} &= -10^{-6}\,, ~\xi_{L2}= - 0.05,~\xi_{L3}=0.02,~ y_{R1}=0.5\times 10^{-4}\,,  \nonumber \\
 y_{R2}&=1\,, ~y_{R3}=0.5\,, ~m_{X(N_k)}=1\, {\rm TeV},~ m_{H^\pm_I}=1.1\, {\rm TeV}\,, \label{eq:benchmark}
 \end{align}
 where these parameter  values have been taken in such a way that  the current upper bounds of the  LFV processes shown in Table~\ref{tab:V_LFV} are satisfied  and  $a_\mu \sim O(10^{-9})$ is achieved.    As a result,  the corresponding values for $BR(\ell_i \to \ell_j \gamma)$, $BR(\ell_i \to 3 \ell_j)$, and $a_\ell$ are obtained as:
  \begin{align}
  \left( BR(\mu \to e \gamma)\,, BR(\tau \to \mu \gamma) \right) & = \left(  4.24\times 10^{-13}\,,~3.47 \times 10^{-8} \right) \,, \nonumber \\
  BR(\mu \to 3e)^{\gamma,\, Z,\,\rm Box} &= \left( 2.62 \times 10^{-16}\,, ~9.09 \times 10^{-28}\,, ~6.37 \times 10^{-35} \right) \,,  \nonumber \\
  BR(\tau \to 3 \mu)^{\gamma,\, Z, \,\rm Box}  & = \left( 8.64 \times 10^{-10}\,,~ 7.09\times 10^{-18}\,, ~1.87 \times 10^{-10} \right)\,, \nonumber \\
  a_{e,\, \mu,\, \tau} & = \left( 9.15\times 10^{-21},\, 9.24 \times 10^{-10}, \, -3.13 \times 10^{-9}\right)\,. \label{eq:test}
  \end{align}
From the simple analysis, it can be clearly seen that in order to obtain  $a_\mu$ of $O(10^{-9})$, the values of the associated parameters have to be  $|\xi_{L2}|\sim O(0.01)$ and $y_{R2}\sim O(1)$. Then,  $\mu\to e \gamma$ inevitably  gives a strict constraint on the $\xi_{L1}$ and $y_{R1}$ parameters. If $y^k_{L1}$  are of the order of $10^{-3}-10^{-2}$, which are the typical magnitudes for explaining  the neutrino data with $\lambda_5\ll 1$~\cite{Ma:2006km}, the result of $\xi_{L1}\sim O(10^{-6})$ or $\xi_{L1} \approx 0$ has to rely on the cancellation in $\sum_k y^k_{L1} h^k_L$. Because  $\xi_{L1}, \, y_{R1}\ll 1$,  the contributions to $\mu \to 3e$ from the Z-penguin and box diagrams are negligible. For $\tau \to 3 \mu$ decay, the $Z$-penguin contribution is still negligible; however, the box-diagram contribution is somewhat larger and is a  factor of 5 smaller than the photon-penguin contribution.   Based on these results, it is sufficient to only consider  the photon-penguin diagram effects when studying the $\mu\to 3 e$ and $\tau \to 3\mu$ decays. In addition, due to $\xi_{L1}\ll 1$, it indicates $y^k_{L1} < y^k_{L2,L3}$. Accordingly, we take the NO case for the neutrino mass matrix in the numerical analysis.

\section{Numerical Analysis}

According to previous analysis, it was found that the neutrino mass, the LFV, and the lepton $g-2$ share some common parameters; however,  the correlated  parameters appearing in the different phenomena have different forms. 
%For instance, the neutrino mass matrix elements depend on $\sum_k y^k_{Li} y^k_{Lj}$ and the $\ell_i \to \ell_j \gamma$, $\ell_i \to 3 
%\ell_j$, and muon $g-2$ processes are associated with $\xi_{Li} = \sum_k y^k_{Li} h^k_L$.  
 From Eq.~(\ref{eq:v_nu_mass}), it is known that we need $\sum_k y^k_{Li'} y^k_{Lj'} \sim 10^{-3}$ and $\sum_k y^{k}_{L1} y^k_{Li}\sim 10^{-4}$ to fit the neutrino mass matrix for the  NO case. From Eq.~(\ref{eq:benchmark}), it is seen that we need $ \xi_{L1} \approx 0$ and $\xi_{L2}\sim O(0.01)$ to satisfy the LFV constraints and to explain the muon $g-2$ excess; that is, different lepton flavor Yukawa couplings $y^k_{Li}$ should be  different in terms of their signs and in sizes.  In order to show that the scotogenic model can accommodate the relevant phenomena in the same parameter spaces, in this section, we numerically demonstrate  that the accommodation can be achieved in the model. 
 
 \subsection{Allowed parameter spaces from the parameter scan}
 
\begin{figure}[phtb]
\includegraphics[scale=0.6]{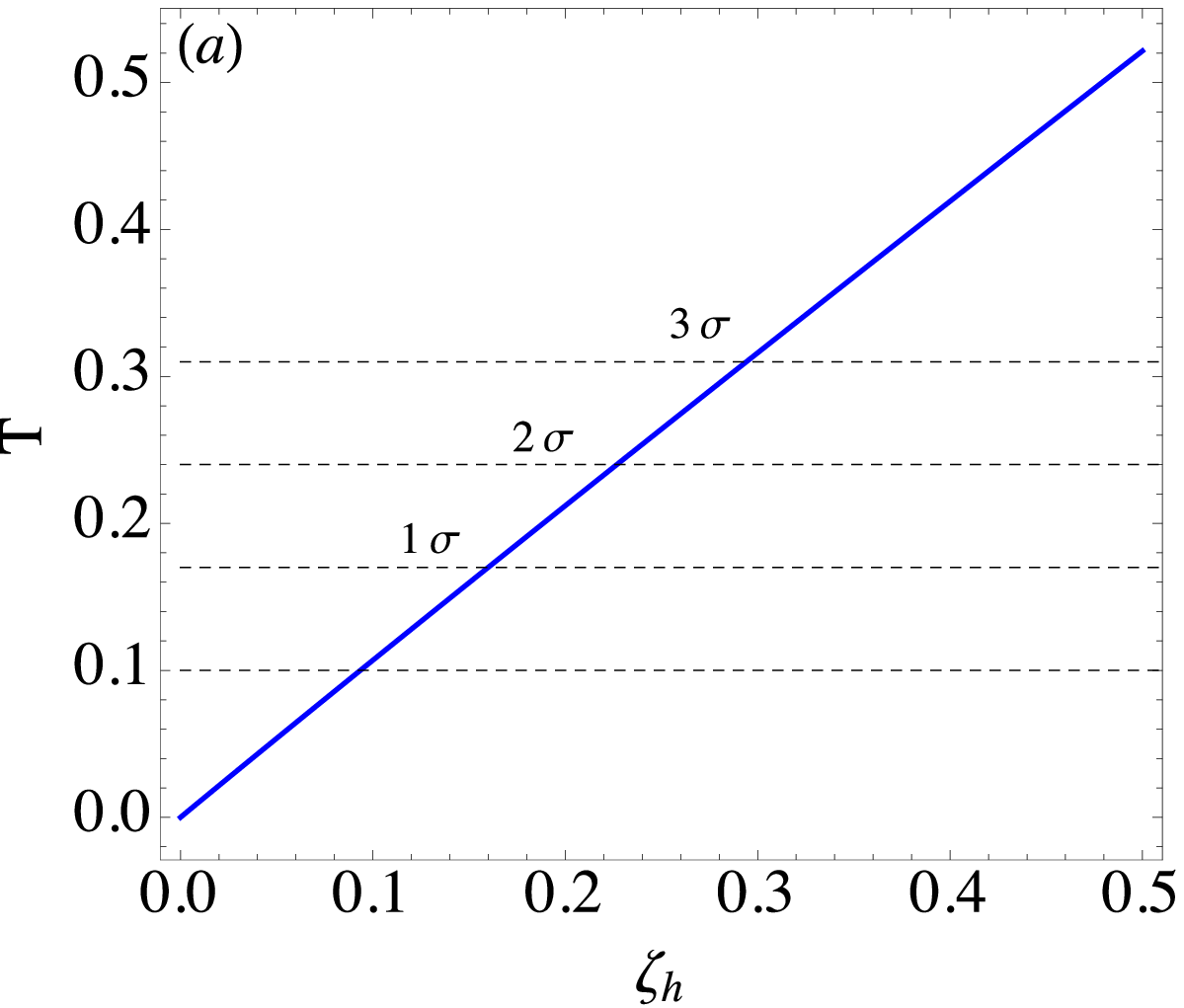}
\includegraphics[scale=0.6]{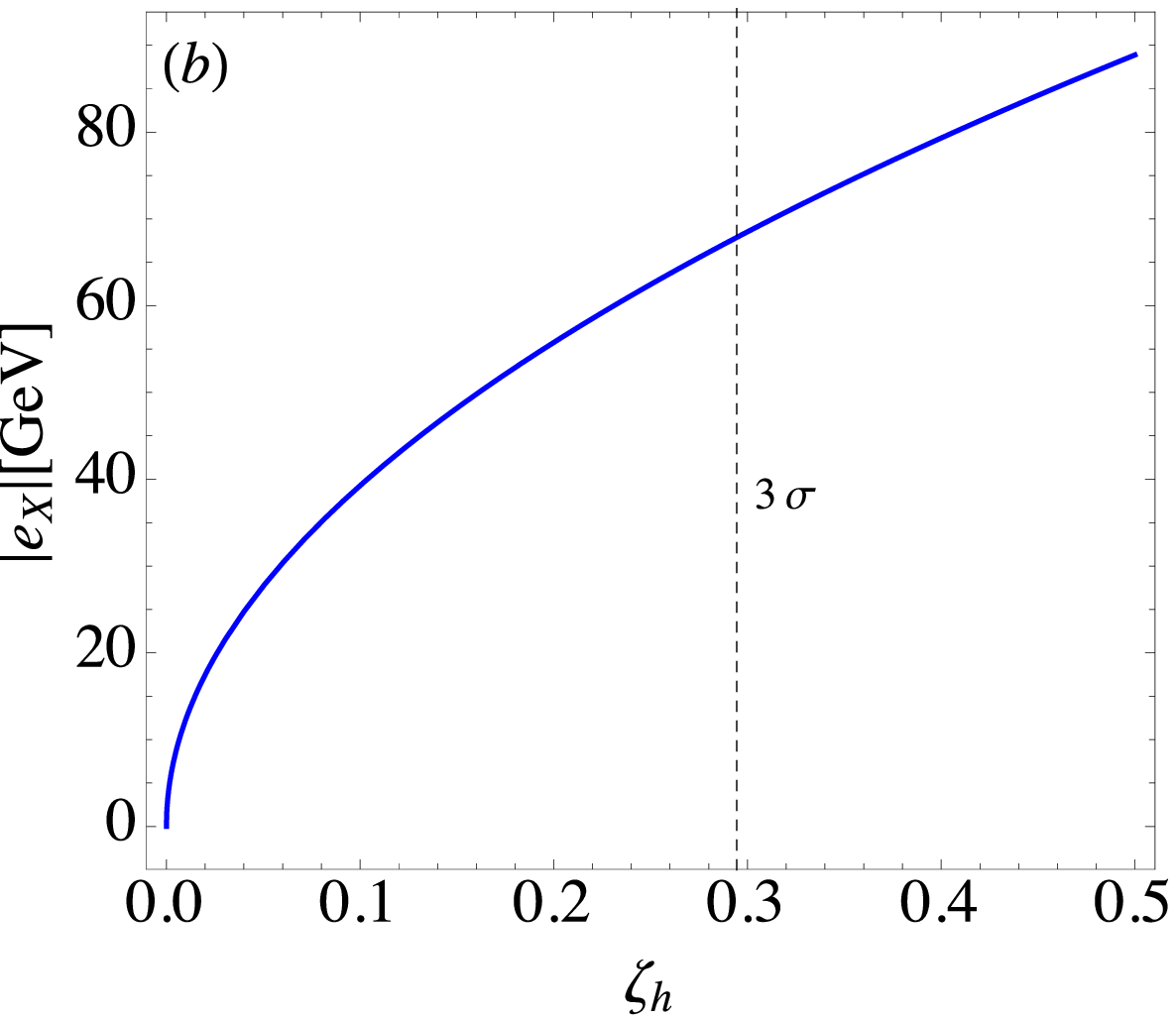}
 \caption{ (a) $T$ parameter as a function of $\zeta_h=\sum_k (h^k_L)^2$, where the dashed lines are the experimental central value with 0, 1, 2, and 3 $\sigma$. (b) Mass splitting ($e_X$) within vector-like lepton doublet as a function of $\zeta_h$, where the vertical dashed line corresponds to the $T$ parameter with $3\sigma$ errors.   }
\label{fig:zeta_h}
\end{figure}

 Since the $\xi_{L i}$ parameters are combined by $h^k_L$ and $y^k_{Li}$,  we first study the limit on $h^k_L$.  As discussed earlier,  the mass splitting in vector-like lepton doublet is $\Delta m_X =|e_X|$, and  the direct bound is from the electroweak oblique parameters $S$ and $T$. Using the results in Eq.~(\ref{eq:ST}), it can be seen that $S$ is far smaller than the current measurement. For instance, with $\zeta_h=1$, we obtain $S\approx 8\times 10^{-3}$; that is, the $S$ parameter cannot constrain the $\zeta_h$ parameter. 
 In order to understand the constraint from the $T$ parameter, we show $T$ as a function of $\zeta_h$ in Fig.~\ref{fig:zeta_h}(a), where the dashed lines denote the experimental central value with 0, 1, 2, and 3$\sigma$ errors, and $m_{X}=m_{N_k}=1$ TeV and the positive sign in $e_X$ are used. From the plot, it can be seen that $T$ linearly depends on $\zeta_h$. If we take $3\sigma$ as the maximum value of $T$, we obtain $\zeta_h < 0.3$.  Moreover, we show $|e_X|$ as a function of $\zeta_h$ in Fig.~\ref{fig:zeta_h}(b), where the vertical dashed line corresponds to the $T$ parameter with $3\sigma$ errors. From the result, it can be found that the maximum value of $|e_X|$ is around $68$ GeV. Our result is consistent with that obtained in~\cite{Arina:2012aj}.  According to the result, when $h^1_L\sim h^2_L\sim h^3_L$, the upper limit of each parameter then is $h^k_L \sim 0.3$. 
 
     %%%%

We next numerically show that $\zeta_h < 0.3$, $\xi_{L1}=0$, $\xi_{L2}\sim O(0.01)$, and that the $m^{\nu}_{ij}$ values shown in Eq.~(\ref{eq:v_nu_mass}) can be
 accommodated in the model. We note that because $\xi_{L1}\ll 1$, we use $\xi_{L1}=0$ in the numerical analysis. From $\xi_{L1}=0$, we can set:
\begin{equation}
y^1_{L1} = - \frac{h^2_L}{h^1_L} y^2_{L1} - \frac{h^3_L}{h^1_L} y^3_{L1}\,.
\end{equation}
To find the allowed parameter spaces, we scan the remaining 11 free parameters in the regions   chosen as:
 \begin{equation}
 h^k_L = [-0.55, 0.55],   ~ y^{k}_{L i} = [-10,10]\times 10^{-2}\,. \label{eq:ranges}
 \end{equation}
In the calculations, we fix $m_{0}=1$ TeV and $m_{S_I}\approx m_{A_I} \approx 1.1$ TeV. In addition,    the value of $|m_{A_I}-m_{S_I}|$  is taken to fit ${\cal M}_0 \approx 6.12 \times 10^{-9}$ GeV. From the $|m^{\nu}_{ij}|$ values shown in Eq.~(\ref{eq:v_nu_mass}), the corresponding  ranges  for $Y_{ij}\equiv |\sum_k y^k_{Li} y^k_{Lj}|$ can then be written  as:
 \begin{align}
Y_{11} & \approx  (1.80 - 7.35) \times 10^{-4}\,, ~ Y_{12}  \approx  (1.96 - 13.39) \times 10^{-4}\,, ~ Y_{13}\approx Y_{12}\,, \nonumber \\
Y_{22} & \approx  (3.92-5.39)\times 10^{-3}\,, ~ Y_{23} \approx  (3.27 - 3.59)\times 10^{-3}\,, ~ Y_{33} \approx  (3.59-5.06)\times 10^{-3}\,.
 \label{eq:Y_ranges}
 \end{align}
 
\begin{figure}[phtb]
\includegraphics[scale=0.5]{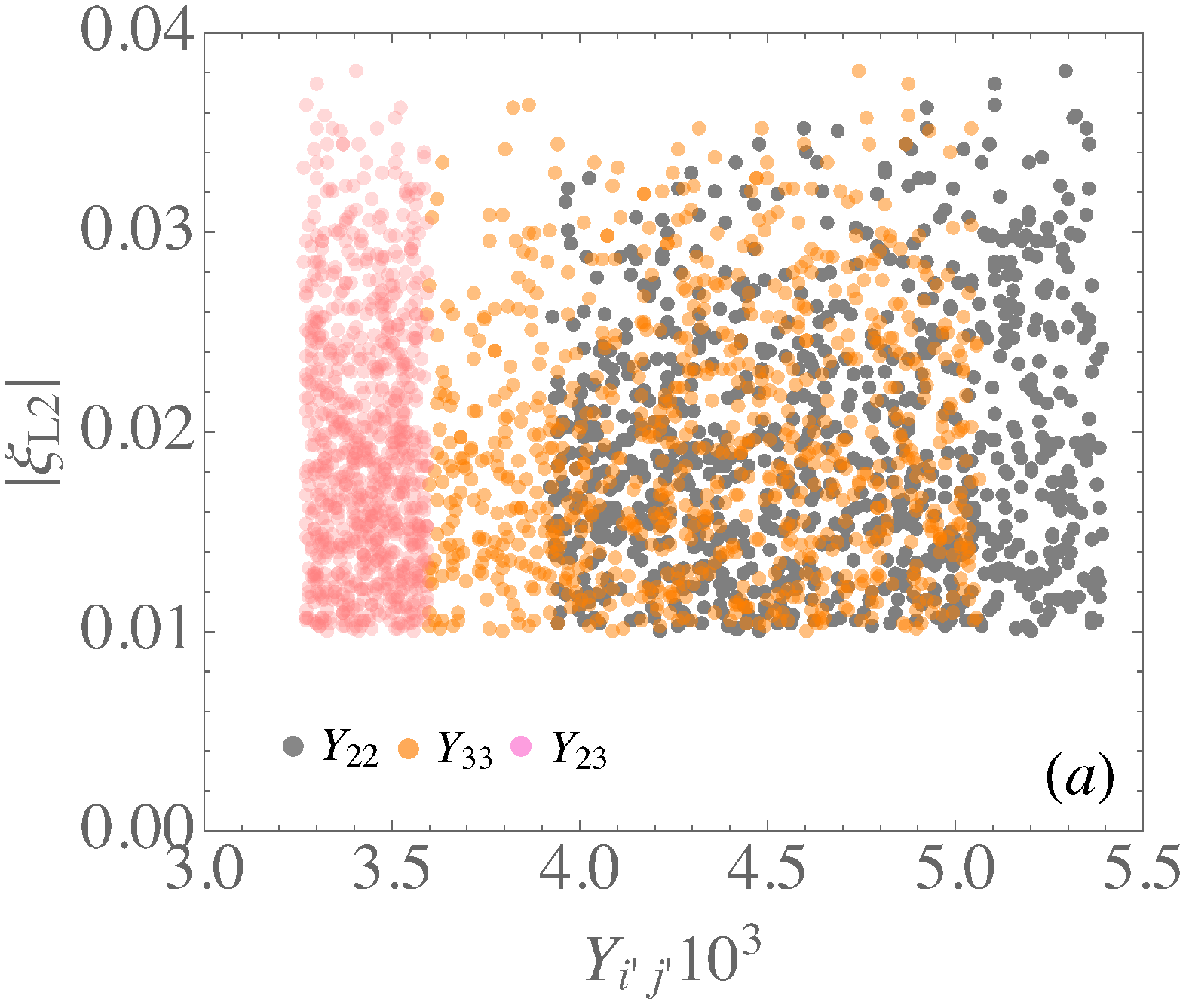}
\includegraphics[scale=0.5]{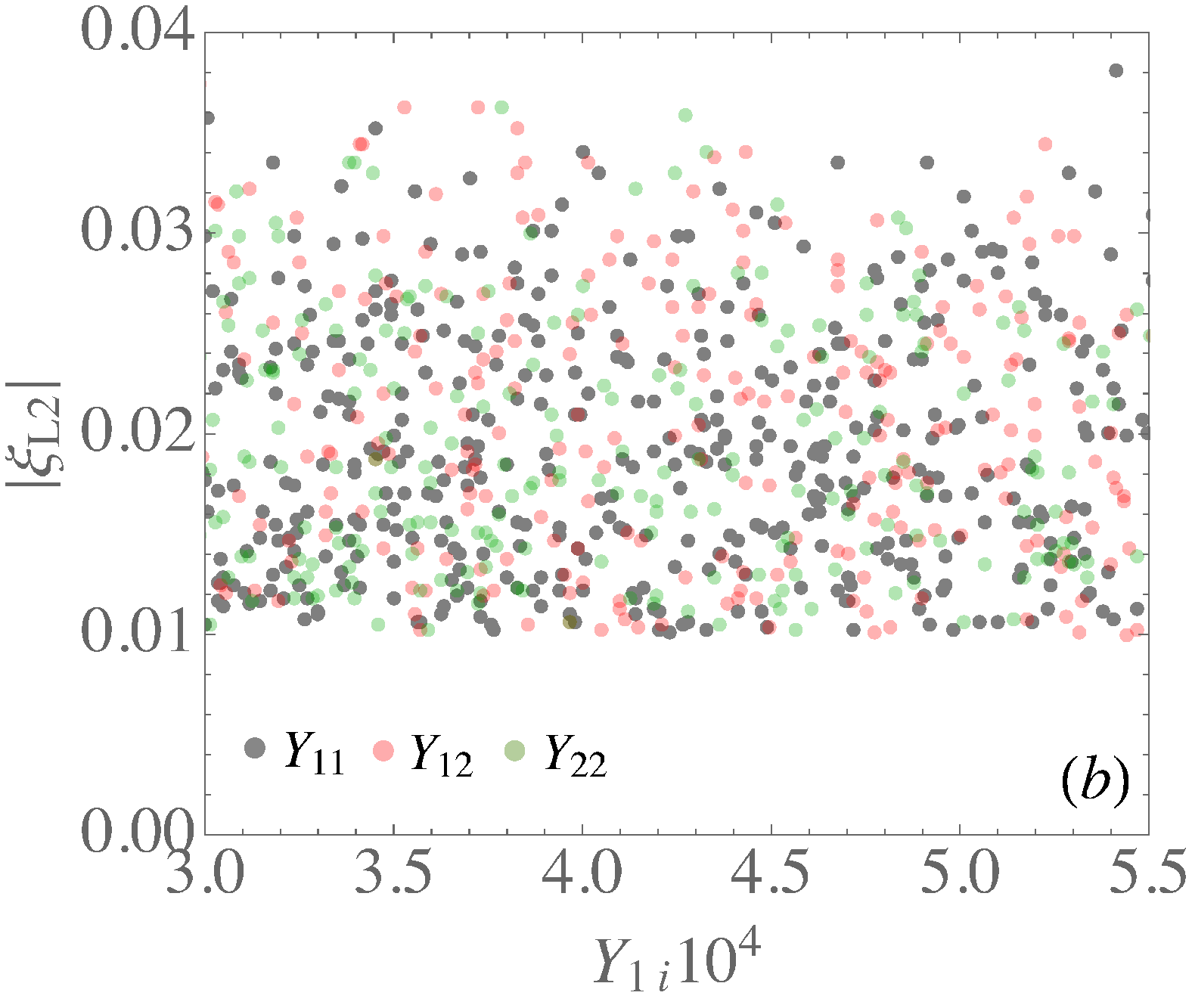}
 \caption{Correlation between $|\xi_{L2}|$ and  $Y_{i'j'}$ ($Y_{1i}$), where $i'(j')=2,3$; $i=1,2,3$, and $\zeta_h<0.3$ and the ranges of $Y_{ij}$ in Eq.~(\ref{eq:Y_ranges}) are satisfied.}
\label{fig:yy_ij}
\end{figure}

 Using $5\times 10^{8}$ random sampling points and the chosen scan ranges in Eq.~(\ref{eq:ranges}),  we show the correlation between $|\xi_{L2}|$ and $Y_{22,33,23}$ in Fig.~\ref{fig:yy_ij}(a), where $|\xi_{L2}|>0.01$, $\zeta_h < 0.3$, and the ranges in Eq.~(\ref{eq:Y_ranges}) are satisfied. The correlation between $|\xi_{L2}|$ and $Y_{1i}$ is shown in Fig.~\ref{fig:yy_ij}(b). From the analysis, it can be seen that $0.01< |\xi_{L2}| < 0.04$ can have a good match with $Y_{ij}$, which are determined by the neutrino data. 
 
     %%%%

\begin{figure}[phtb]
\includegraphics[scale=0.4]{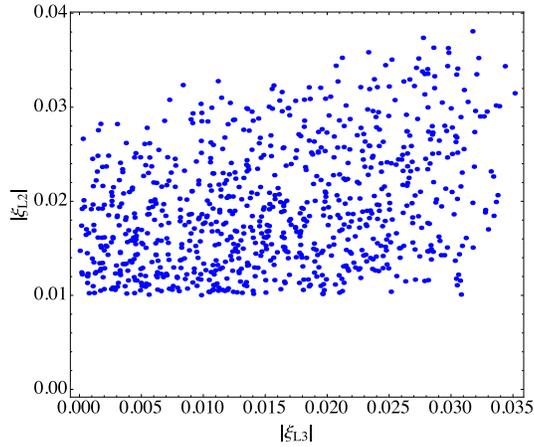}
 \caption{Correlation between $|\xi_{L2}|$ and $|\xi_{L3}|$}
\label{fig:xi2_xi3}
\end{figure}

As mentioned earlier, $\mu \to e \gamma$ gives a strong constraint on the $\xi_{L1}$ and $y_{R1}$ parameters; therefore, a simple way to comply with the requirement is to take $\xi_{L1}=0$ and $y_{R1}=0$. However, even so, the $\tau \to \mu \gamma$ decay may play an important role in  constraining  the parameters, where from Eqs.~(\ref{eq:LFV_as}) and (\ref{eq:LFV_bs}),  the related parameters are $y^*_{R2} y_{R3}$, $y^*_{R2} \xi_{L3}$, and $\xi^*_{L2} y_{R3}$.  If we take the limit with $y_{R3}=0$, the BR for $\tau\to \mu \gamma$ does not vanish due to the  $y^*_{R2} \xi_{L3}$ effect. Since $\xi_{L3}$ is a combination of $h^k_L$ and $y^{k}_{L3}$, which are correlated with $\xi_{L2}$, the $T$ parameter, and the neutrino mass matrix, we cannot arbitrarily tune $\xi_{L3}$ to be small.  In order to see if $\xi_{L3}$ can be small when the oblique $T$ parameter and  neutrino data are satisfied, we show the correlation between $|\xi_{L2}|$ and $|\xi_{L3}|$ in Fig.~\ref{fig:xi2_xi3}, where the conditions used to determine the parameter values  are the same as those shown in Fig.~\ref{fig:yy_ij}. From the result, it can be clearly seen that when $|\xi_{L3}| < 0.01$, $|\xi_{L2}|$ can still reach 0.03. Hence, $BR(\tau \to \mu \gamma)$ can be well controlled in the model. 

     %%%%

 \subsection{$\mu \to e \gamma$ and $\mu\to 3 e$}

\begin{figure}[phtb]
\includegraphics[scale=0.6]{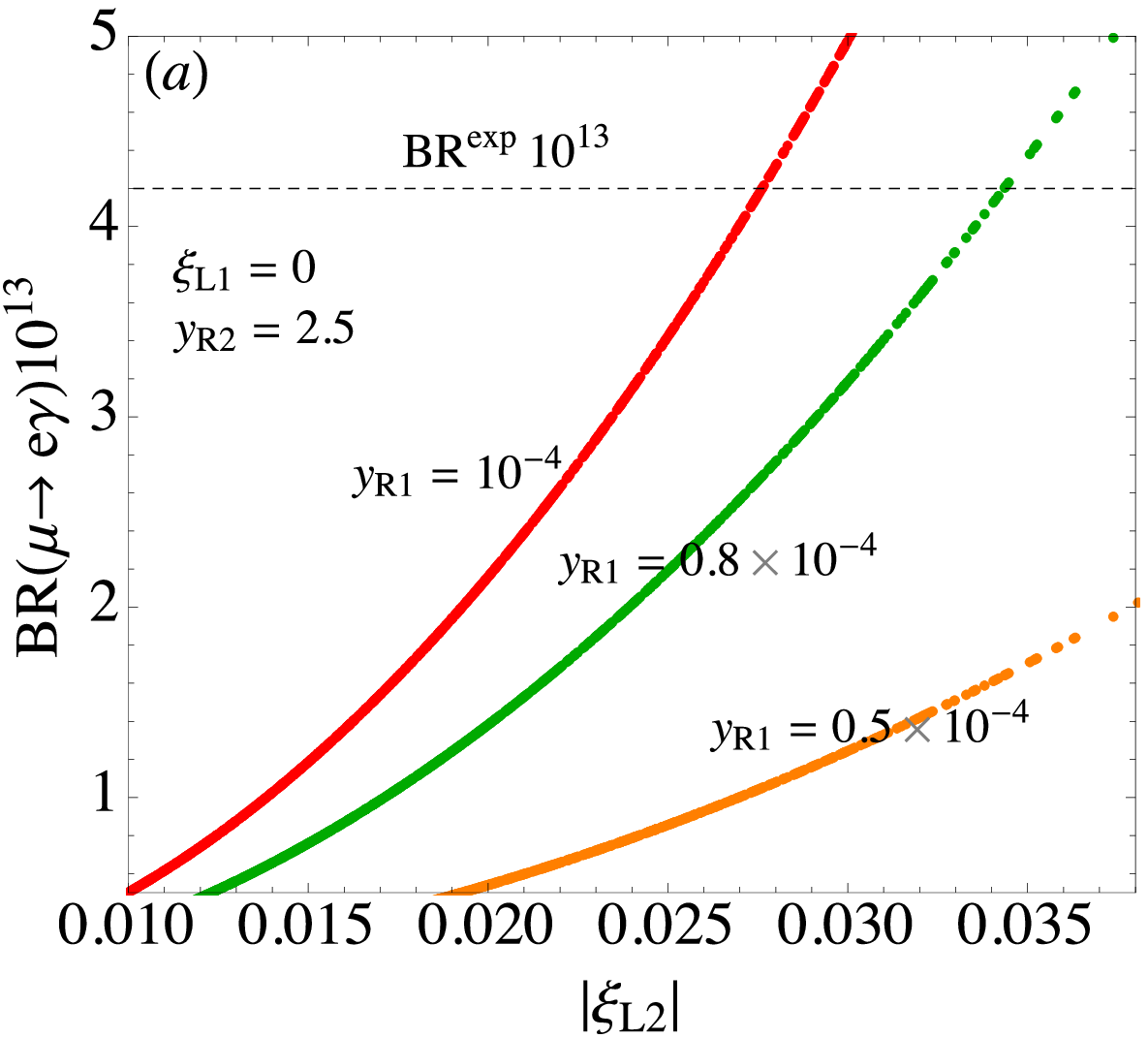}
\includegraphics[scale=0.6]{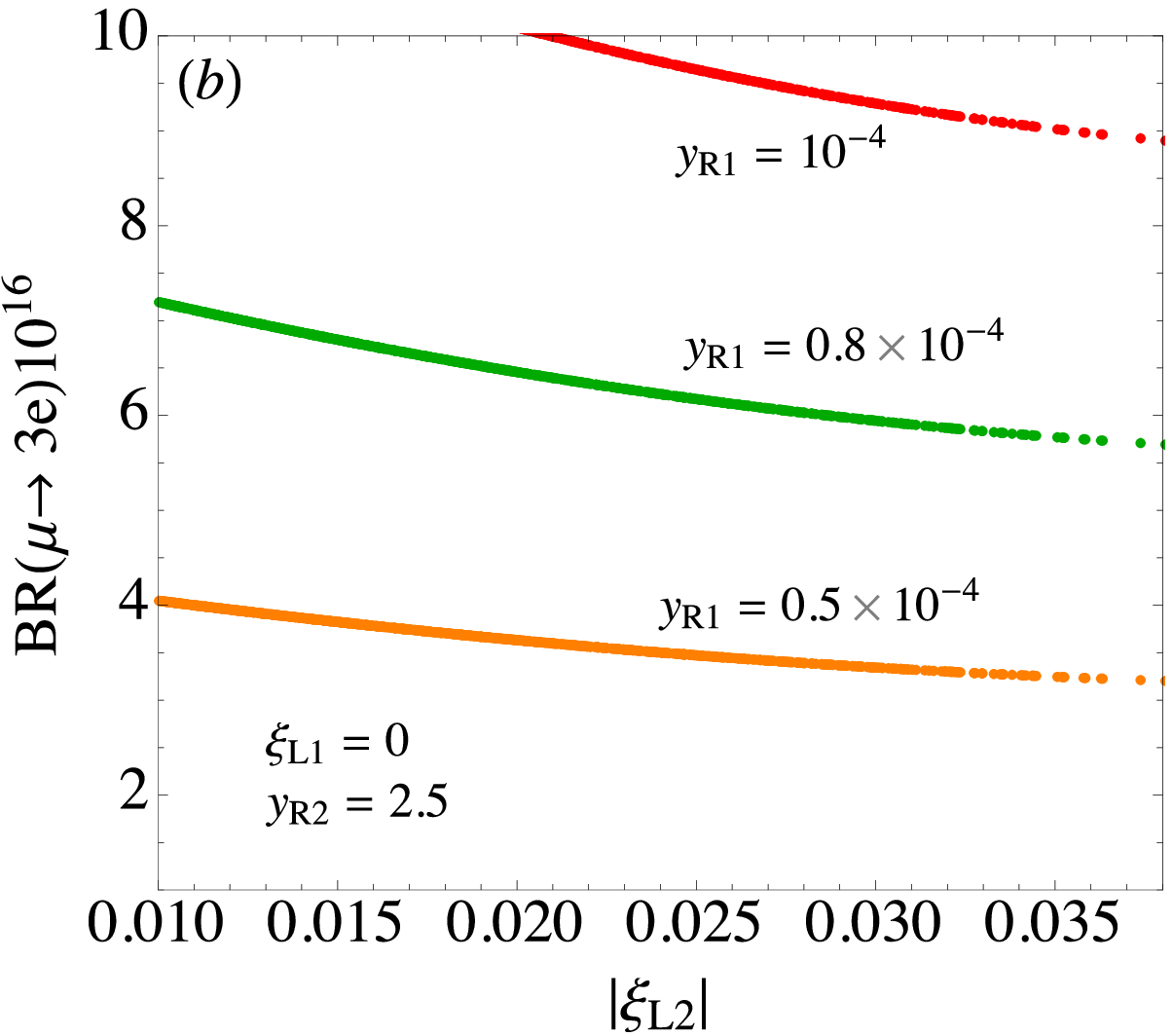}
 \caption{  (a) scatters of $BR(\mu\to e \gamma)$ (in units of $10^{-13})$ as a dependence of $|\xi_{L2}|$, where the allowed data points are applied, $\xi_{L1}=0$ and $y_{R2}=2.5$ are fixed, and the results with $y_{R1}=(1.0,\; 0.8,\; 0.5)\times 10^{-4}$ are shown. (b) scatters of $BR(\mu\to 3e)$ (in units of $10^{-16}$), where the same conditions used in (a) are applied.  }
\label{fig:LFV_mu}
\end{figure}

 According to the indication shown in Eq.~(\ref{eq:test}),  we have to take a small $\xi_{L1}=0$ to fit the  BR for $\mu \to e \gamma$.  For simplicity, $\xi_{L1}=0$ is fixed in the parameter scan. From Eqs.~(\ref{eq:LFV_as}) and (\ref{eq:LFV_bs}), it can be seen that even when using $\xi_{L1}=0$, the BR for $\mu\to e \gamma$ is still dictated by $y^*_{R1} \xi_{L2}$; that is,  the $\mu \to e \gamma$ also gives a strict constraint on the $y_{R1}$ parameter. In order to understand how $BR(\mu\to e \gamma)$ is sensitive to $\xi_{L1}$, $BR(\mu\to e \gamma)$ (in units of $10^{-13}$) as a function of $|\xi_{L2}|$ is shown in Fig.~\ref{fig:LFV_mu}(a), where  the allowed parameter spaces   are applied; $y_{R2}=2.5$ is used,  and the results for $y_{R1}=(1.0,\;0.8,\;0.5)\times 10^{-4}$ are shown, respectively, in the plot.  It can be seen that $\mu \to e \gamma$ can further exclude some parameter regions if the $y_{R1}$ values approach $10^{-3}$ from below. We also show the dependence for the $\mu \to 3 e$ decay in Fig.~\ref{fig:LFV_mu}(b); however, the result is far less than the current upper limit.

     %%%%

\subsection{$\tau \to \mu \gamma$ and $\tau \to 3\mu$}

\begin{figure}[phtb]
\includegraphics[scale=0.6]{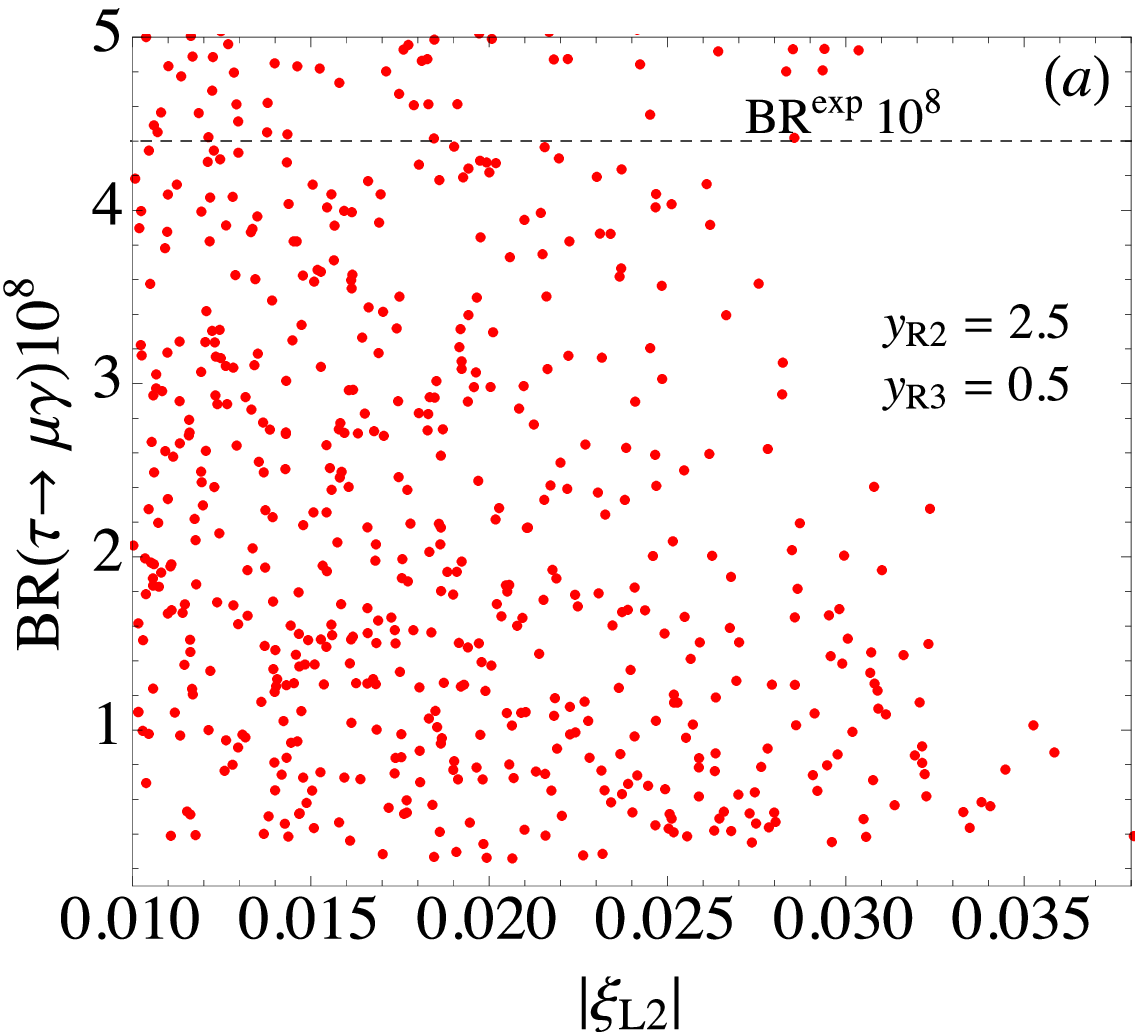}
\includegraphics[scale=0.6]{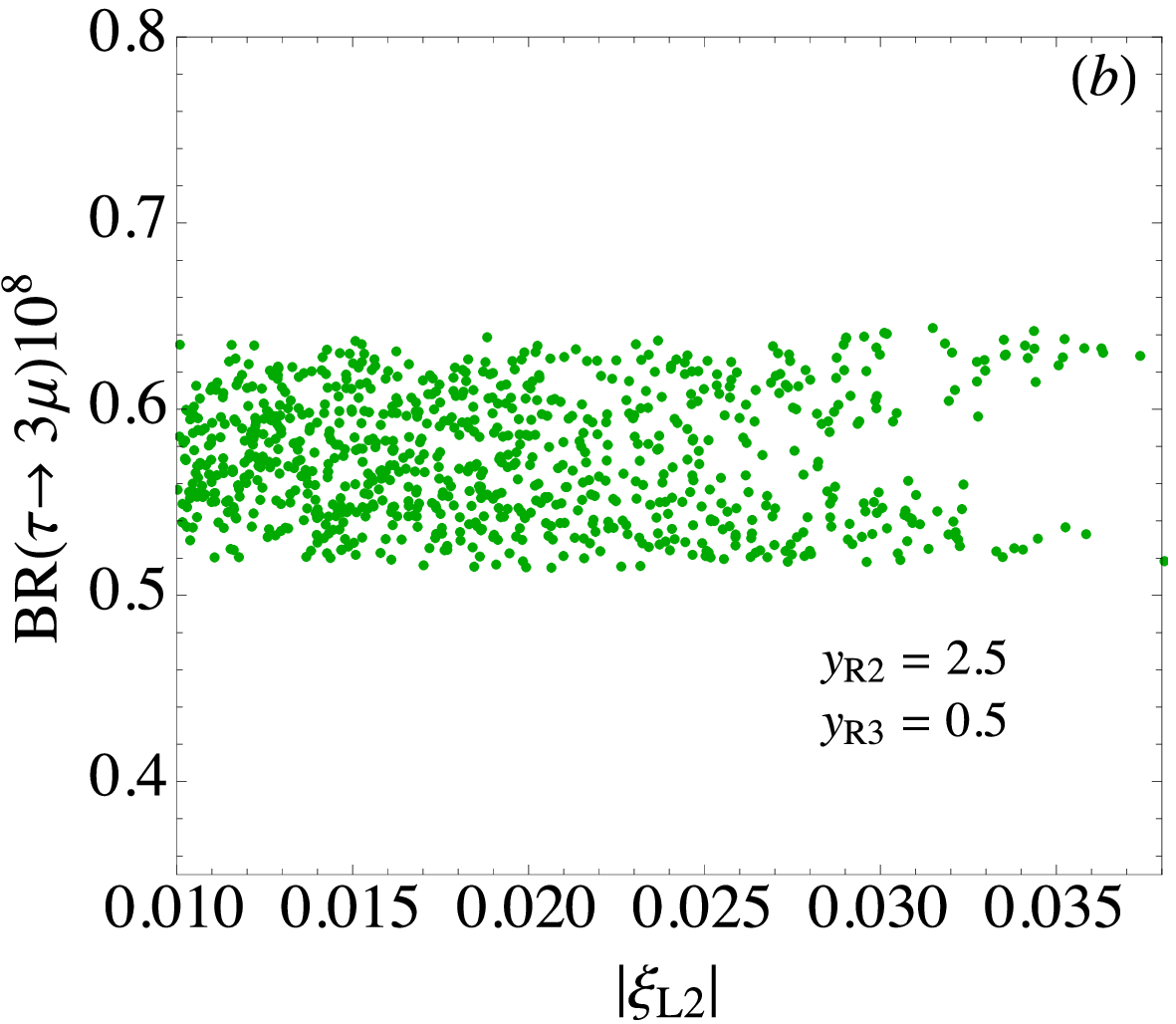}
 \caption{  Scatters for (a) $BR(\tau \to \mu \gamma)$ and (b) $BR(\tau \to 3 \mu)$, where $y_{R2}=2.5$ and $y_{R3}=0.5$ are fixed. }
\label{fig:LFV_tau}
\end{figure}

Using the allowed data points, which are obtained by the parameter scan, we show the scatters of $BR(\tau \to \mu \gamma)$ ( in units of $10^{-8}$) with respect to $|\xi_{L2}|$  in~\ref{fig:LFV_tau}(a), where $y_{R2}=2.5$ and $y_{R3}=0.5$ are used, and the horizontal dashed line is the experimental upper limit. It can be seen that $BR(\tau \to \mu \gamma)$ indeed can further bound the parameter.  In 
order to retain $a_{\mu}\sim 10^{-9}$,  we can take a smaller value for $y_{R3}$.  We also show the scatter plot for  $BR(\tau \to 3 \mu)$ in units of $10^{-8}$ in~\ref{fig:LFV_tau}(b). Although  the resulting $BR(\tau \to 3 \mu)$ is still smaller than the current upper limit by one oder of magnitude,  the allowed region can still reach the Belle II sensitivity of tau physics.  
     %%%%

\begin{figure}[phtb]
\includegraphics[scale=0.6]{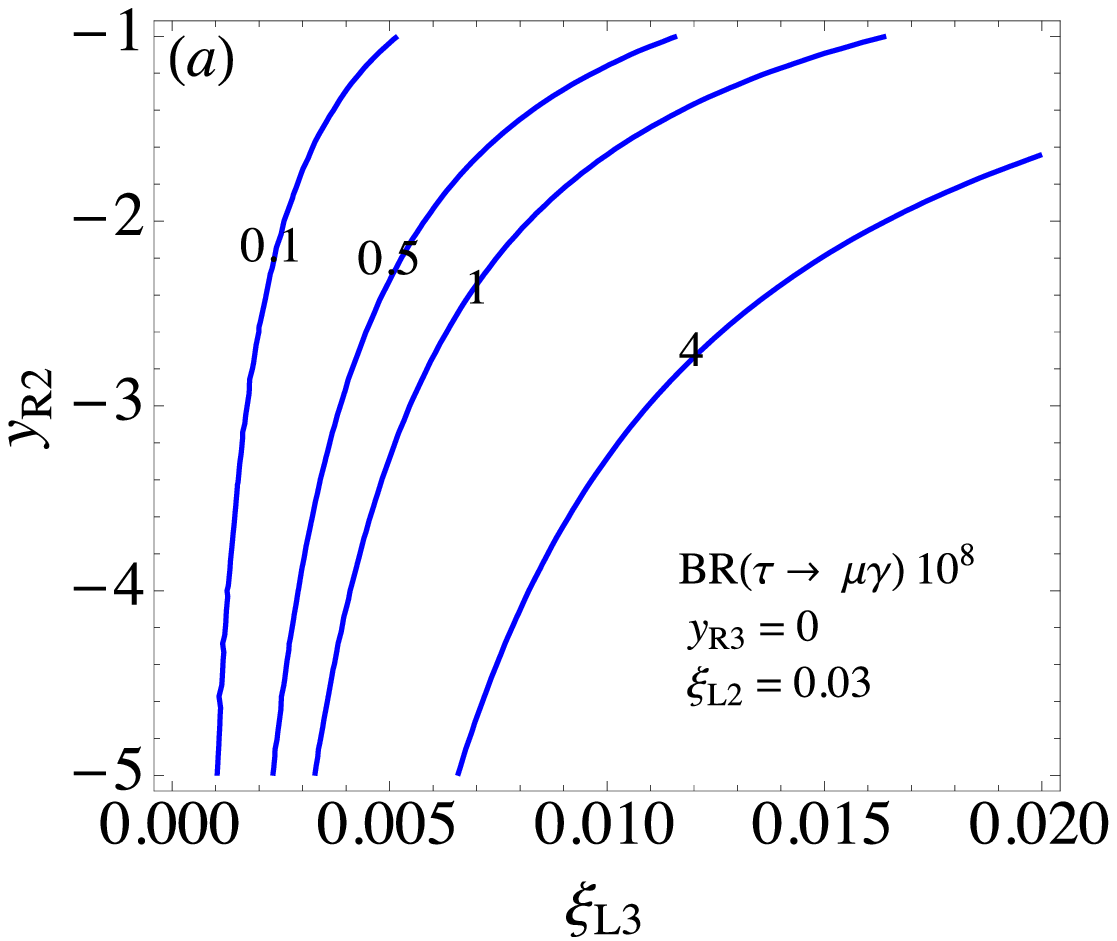}
\includegraphics[scale=0.6]{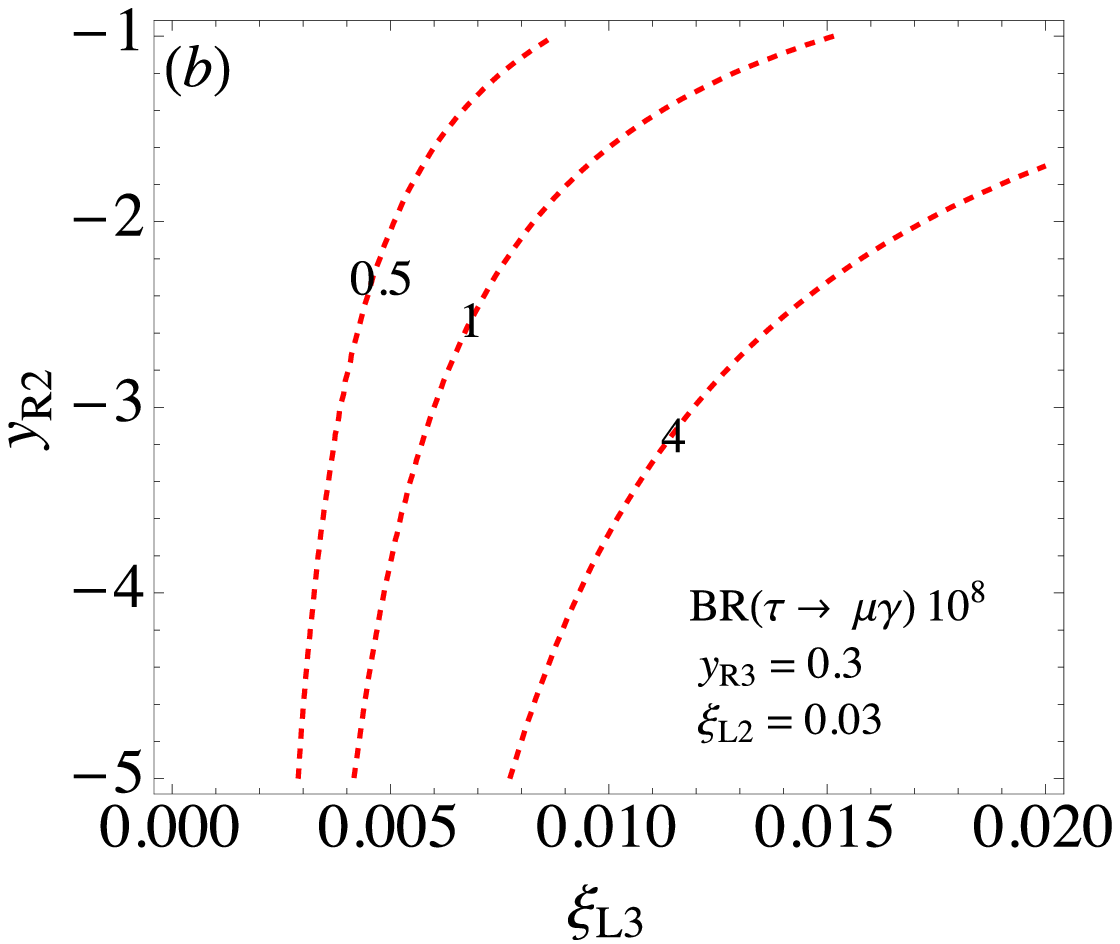}
 \caption{ Contours for $BR(\tau \to \mu \gamma)$ (in units of $10^{-8}$) as a function of $\xi_{L3}$ and $y_{R2}$, where we fix $y_{R3}=0$ in (a) and $y_{R3}=0.3$ in (b), respectively. In both plots,  $\xi_{L2}=0.03$ is used.  }
\label{fig:Brtaumuga}
\end{figure}

As mentioned before, $BR(\tau \to \mu \gamma)$ depends on the $y^*_{R2} y_{R3}$, $y^*_{R2} \xi_{L3}$, and $\xi^*_{L2} y_{R3}$ parameters. To gain a better understanding of   the correlations among parameters, we show the contours of $BR(\tau \to \mu \gamma)$ (in units of $10^{-8})$ as a function of $\xi_{L3}$ and $y_{R2}$ in Fig.~\ref{fig:Brtaumuga}, where  plot (a) and plot (b) denote $y_{R3}=0$ and $y_{R3}=0.3$, respectively, and $\xi_{L2}=0.03$ is fixed in both plots. From the results, it can be seen that $BR(\tau \to \mu \gamma)$ does not vanish at $\xi_{L3}=0$ when $y_{R3} \neq 0$. If  the Belle II experiment does not find any event for the $\tau \to \mu \gamma$ decay at the sensitivity  of $10^{-9}$~\cite{Aushev:2010bq}, a simple way to suppress the BR for $\tau \to \mu \gamma$ in the model is to take $y_{R3}=0$.

     %%%%

\subsection{muon $g-2$}

\begin{figure}[phtb]
\includegraphics[scale=0.5]{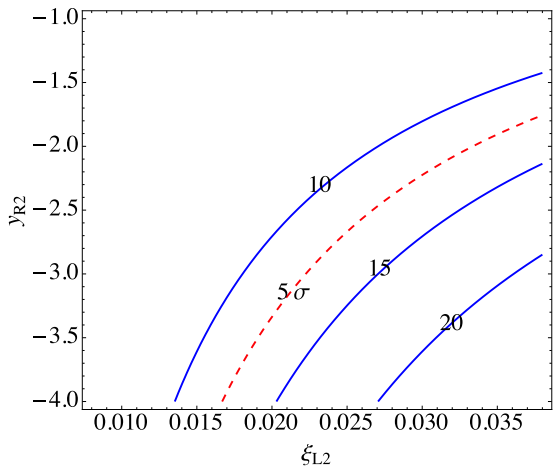}
 \caption{Muon $g-2$ (in units of $10^{-10}$) as a function of $\xi_{L2}$ and $y_{R2}$, where the dashed line denotes the $5\sigma$ result when the experimental and theoretical uncertainties are reduced by a factor of 4 and 2, respectively. }
\label{fig:muon_gm2}
\end{figure}

 According to above analysis, it is known that  the range of $0.01< |\xi_{L2}|<0.04$ is allowed in the  model. Although we only show the positive values for $\xi_{Li}$, indeed, the same allowed region is also suitable for the negative $\xi_{Li}$ with the exception of sign. From Eq.~(\ref{eq:lepton_gm2}), it can be seen that $\xi_{L2}$ and $y_{R2}$ have to be opposite in sign in order to get a positive $a_\mu$. To see the influence of inert charged-Higgs effects on the muon $g-2$, we show $a_{\mu}$ (in units of $10^{-10}$) as a function of positive $\xi_{L2}$ and negative $y_{R2}$ in Fig.~\ref{fig:muon_gm2}, where the dashed line denotes the $5\sigma$ result when the experimental and theoretical uncertainties are reduced by a factor of 4 and 2, respectively. The same result can be applied to the negative $\xi_{L2}$ and positive $y_{R2}$. It can be concluded that $a_{\mu} \sim O(10^{-9})$ can be realized in the model when the experimental constraints are included. 
 
     %%%%

\section{Summary}

Based on the scotogenic model proposed in~\cite{Ma:2006km}, we extend the model by including an $Z_{2}$-odd vector-like lepton doublet ($X$) in order to resolve the muon $g-2$ excess through the mediation of inert charged-Higgs. 

In the model, two new Yukawa interactions, i.e. $\bar X_L H_I \ell_R$ and $\bar X_L \tilde{H} N_k$, play the main key effects.  In addition to the new Yukawa couplings,  the induced muon $g-2$ also depends on other Yukawa couplings, which are determined by the neutrino mass matrix elements of the order of   $10^{-3}-10^{-2}$ eV. It was found that the case with $m_{X, N_k} > m_{S_I(A_I)}$ cannot significantly enhance the muon $g-2$ because of the bound from the direct dark matter detection.  Thus, the suitable dark matter candidate in the model is the lightest $Z_2$-odd Majorana lepton.

Lepton-flavor violation processes, especially $\mu \to e \gamma$ and $\tau \to \mu \gamma$, make strict constraints on the relevant parameters. Nevertheless, we found that the resulting muon $g-2$ can reach $O(10^{-9})$ when the 11 independent parameter values satisfy  the experimental measurements, such as lepton-flavor violation, neutrino oscillations, and electroweak oblique parameters. Moreover, the branching ratio for $\tau \to \mu \gamma$ can be well controlled and can reach the sensitivity of Bell II with an integrated luminosity of  50 ab$^{-1}$. \\

\section*{Acknowledgments}

This work was partially supported by the Ministry of Science and Technology of Taiwan,  
under grants MOST-106-2112-M-006-010-MY2 (CHC).

\end{document}